\documentclass[a4paper]{iopart}

\usepackage{cite}
\usepackage[dvips]{graphicx}
\usepackage{times}
\usepackage{latexsym}
\usepackage{amssymb}
\usepackage{color}

\newcommand{\cf}{\textit{cf.}}
\newcommand{\ie}{\textit{i.e.}~}

\begin{document}

\title[Critical Phenomena in Neutron Stars II: Head-on Collisions]{Critical
Phenomena in Neutron Stars II: Head-on Collisions}

\author{Thorsten Kellerman$^{1}$, Luciano Rezzolla$^{1,2}$ and David  Radice$^{1}$}

\address{$^1$ Max-Planck-Institut f\"ur Gravitationsphysik, Albert Einstein
Institut, Potsdam, Germany}

\address{$^2$ Department of Physics and Astronomy, Louisiana State
  University, Baton Rouge, USA}

\begin{abstract}
We consider the head-on collision of equal-mass neutron stars boosted
towards each other and we study the behavior of such systems near the
threshold of black-hole formation. In particular, we confirm the
previous findings by~\cite{Jin:07a} that a type-I critical phenomenon
can be observed by fine-tuning the initial mass of the two neutron
stars. At the same time, we argue against the interpretation that the
critical solution is not a perturbed spherical star and show instead
that the metastable star corresponds to a (perturbed) equilibrium
solution on the unstable branch of the equilibrium configurations. As
a result, the head-on collision of two neutron stars near the critical
threshold can be seen as a transition in the space of configurations
from an initial stable solution over to a critical metastable one
which can either migrate to a stable solution or collapse to a black
hole. The critical exponent for this process shows a fine structure
which was already observed in the case of the critical collapse of
scalar fields but never before for perfect fluids.
\end{abstract}

\pacs{
04.25.Dm, 
04.40.Dg, 
04.70.Bw, 
95.30.Lz, 
97.60.Jd
}

\section{Motivation and introduction}
In 1993, Choptuik~\cite{Choptuik93} considered one-parameter families
of solutions, $\mathcal{S}[P]$, of the Einstein-Klein-Gordon equations
for a massless scalar field in spherical symmetry, such that for every
$P > P^\star$, $\mathcal{S}[P]$ contains a black hole and for every $P
< P^\star$, $\mathcal{S}[P]$ is a solution not containing
singularities. He studied numerically the behavior of $\mathcal{S}[P]$
as $P\to P^\star$ and found that the critical solution,
$\mathcal{S}[P^\star]$, is \emph{universal}, in the sense that it is
approached by all nearly-critical solutions regardless of the
particular family of initial data considered. He also found that
$\mathcal{S}[P]$ exhibit discrete self-similarity and that, for
supercritical solutions $(P > P^\star)$, the mass of the black hole
satisfies $M_{\mathrm{BH}} = c | P - P^\star |^\gamma$, with $\gamma$
being an universal constant, \ie not depending on the particular
family of initial data.

After Choptuik's seminal work, similar transitions were discovered for
a wide range of systems, including massive scalar fields and
ultra-relativistic fluids, see~\cite{gundlach_2007_cpg} for a recent
review.  All these phenomena have the common property that, as $P$
approaches $P^\star$, $\mathcal{S}[P]$ approaches a universal solution
$\mathcal{S}[P^\star]$ and that all the physical quantities of
$\mathcal{S}[P]$ depend only on $|P-P^\star|$. In analogy with
critical phase transitions in statistical mechanics, these transitions
in gravitational collapse were later classified as ``type-I'' critical
phenomena, with static or periodic critical solutions and
discontinuous transitions in the vicinity of the critical point, or
``type-II'' critical phenomena, with self-similar critical solutions
and continuous transitions in the vicinity of the critical solution
~\cite{gundlach_2007_cpg}.

The study of critical phenomena in neutron-star (NS) collapse started
with the work by~\cite{evans_1994_cps} on radiation fluids and was
later extended to more general ultra-relativistic equations of state
(EOS)~\cite{Hara96a, brady_2002_bht} and ideal-gas
EOS~\cite{novak_2001_vic, noble_2003_nsr, Noble08a}. In all these
studies the collapse was triggered using strong perturbations and a
type-II critical phenomena was found. Type-I critical phenomena in the
collapse of unstable configuration under very small perturbations was
instead studied only very recently~\cite{liebling_2010_emr,
  Radice:10}.

The first study of critical phenomena the head-on collision of NSs was
instead carried out by Jin and Suen in 2007~\cite{Jin:07a}. In
particular, they considered a series of families of equal mass NSs,
modeled with an ideal-gas EOS, boosted towards each other and varied
the mass of the stars, their separation, velocity and the polytropic
index in the EOS. In this way they could observe a critical phenomenon
of type I near the threshold of black-hole formation, with the
putative solution being a nonlinearly oscillating star. In a
successive work~\cite{wan_2008_das}, they performed similar
simulations but considering the head-on collision of Gaussian
distributions of matter. Also in this case they found the appearance
of type-I critical behaviour, but also performed a perturbative
analysis of the initial distributions of matter and of the merged
object. Because of the considerable difference found in the
eigenfrequencies in the two cases, they concluded that the critical
solution does not represent a system near equilibrium and in
particular not a perturbed Tolmann-Oppenheimer-Volkoff (TOV)
solution~\cite{Jin:07a}.

In this paper we study the dynamics of the head-on collision of two
equal-mass NSs using a setup which is as similar as possible to the
one considered in~\cite{Jin:07a}. While we confirm that the merged
object exhibits a type-I critical behaviour, we also argue against the
conclusion that the critical solution cannot be described in terms of
equilibrium solution. Indeed, we show that, in analogy with what found
in~\cite{Radice:10}, the critical solution is effectively a perturbed
unstable solution of the TOV equations. Our analysis also considers
fine-structure of the scaling relation of type-I critical phenomena
and we show that it exhibits oscillations in a similar way to the one
studied in the context of scalar-field critical
collapse~\cite{Gundlach97f, Hod97}.

The remainder of this paper is organized as follows. In section
\ref{sec:setup} we describe the numerical settings of the simulations,
the properties of the used initial data and we define some of the
quantities that we analysed. In section \ref{sec:results} we show in
details our results, while section \ref{sec:conclusions} is dedicated
to conclusions and discussion. Finally, presented in Appendix A is a
study of the compactness of the metastable solution and a comparison
with its ``hoop radius''. The work reported is closely related to the
analysis carried in a companion paper (hereafter paper I) about
critical phenomena in linearly unstable nonrotating NS
models~\cite{Radice:10}. Because of the logical affinity between the
two works we will often refer the reader to the results obtained
in~\cite{Radice:10}. We use a spacelike signature $(-,+,+,+)$, with
Greek indices running from 0 to 3, Latin indices from 1 to 3 and the
standard convention for the summation over repeated indices. Unless
explicitly stated, all the quantities are expressed in the system of
dimensionless units in which $c=G=M_\odot=1$.
\section{Numerical setup and initial models}\label{sec:setup}
\label{Num_methods}

In what follows we briefly describe the numerical setup used in the
simulations and the procedure followed in the construction of the
initial data.

\subsection{Numerical Setup}
The numerical setup used in our simulations is identical to the one
presented in \cite{Radice:10}. In particular, we use the axisymmetric
code \texttt{Whisky2D}, which has been described in detail
in~\cite{Kellermann:08a} (but see also the more extended discussion in
paper I) and is based on the 3-dimensional code
\texttt{Whisky}~\cite{Baiotti04,Giacomazzo:2007ti,Baiotti08}, to solve
in 2 spatial dimensions the full set of Einstein and
relativistic-hydrodynamics equations. The axisymmetry of the spacetime
is imposed exploiting the ``cartoon'' technique introduced
in~\cite{Alcubierre99a}, while the hydrodynamics equations (which are
written in first-order form and thus do not require double spatial
derivatives) are written explicitly in cylindrical coordinates
(see~\cite{Kellermann:08a} to read about the advantages of this
formulation).

All the simulations presented here have been performed using an
ideal-gas EOS, $p=(\Gamma-1)\rho\epsilon$, where $\rho$ is the
baryonic density and $\epsilon$ the specific internal energy in the
rest frame of the fluid\footnote{We recall that an ideal-gas EOS and a
  polytropic one are identical for fluids undergoing isoentropic
  evolutions.}. The discretization of the spacetime evolution equations
is done using fourth-order finite-differencing schemes, while we
high-resolution shock-capturing (HRSC) methods with a
piecewise-parabolic method (PPM) reconstruction for the hydrodynamics
equations. The time-stepping is done with a third-order
total-variation diminishing (TVD) Runge-Kutta scheme and despite the
use of higher-order methods, the convergence order drops to about 1
after the two NSs have merged and large shocks develop
(see~\cite{Baiotti:2009gk} for a discussion on the convergence order
in relativistic-hydrodynamics simulations).

The spatial discretization is done a grid with uniform resolution,
which we have taken to be either $h=0.1\ M_\odot$ or
$h=0.08\ M_\odot$. Furthermore, as long as nonspinning NSs are
considered, the head-on collision also possesses a symmetry across the
plane midway between the two stars and orthogonal to the colliding
direction. As a result, the problem needs to be solved only for one
star and suitable boundary conditions be applied across the symmetry
plane. The outer boundary of the computational domain is set at
$60\ M_\odot$ and thus rather close to the two stars. However, this is
adequate since we are not interested here in extracting gravitational
waveforms and since we have verified that the violation of the
constraints at the outer boundaries are not larger than elsewhere in
the computational domain.

As a final remark we note that we use the same gauges as those employed
in~\cite{Baiotti06} and thus the slicing is sufficiently
``singularity-avoiding'' that it is not necessary to perform an
excision of the field variables when following the evolution of a
supercritical solution. However, because of the very high-resolution
used, the rapid growth of the rest-mass density is not compensated by
the intrinsic numerical dissipation as instead happens
in~\cite{Baiotti06} or in~\cite{Baiotti08}. As result we excise the
solution of the hydrodynamical quantities only as discussed
in~\cite{Hawke04,Giacomazzo:2007ti} in order to obtain a stable,
long-term solution.

\subsection{Initial data}

Our initial data consists of two equal-mass nonrotating NSs having
initial coordinate separation, computed as the coordinate distance
between the two stellar centres, of
$20\ M_\odot$. Following~\cite{Jin:07a}, we construct these stars
using a polytropic EOS, $p=K\rho^{\Gamma}$, with adiabatic exponent
$\Gamma=2$ and polytropic constant $K=80$, which is equivalent to $K =
0.00298\ c^2 / \rho_{\mathrm{n}}$, and where $\rho_{\mathrm{n}} = 2.3
\times 10^{14} \ \mathrm{g/cm}^3$ is approximately the nuclear
density. The stars in the neighbourhood of the critical solution have
a radius of $R=14.7$ km, a baryon/gravitational (ADM) mass of
$0.760/0.732\,M_{\odot}$. The main properties of the critical solution
are reported in table~\ref{tab:models}, where they are indicated as
model ``A''.

Note that the maximum baryon/gravitational mass for the chosen value
of the polytropic constant is $1.609/1.464\,M_{\odot}$ and thus the
object produced by the collision will have a baryon/gravitational mass
which is above such maximum mass. However, as we will discuss below
and in contrast with the claim made in~\cite{Jin:07a}, this can still
lead to equilibrium solution for a TOV star.

\begin{table}[ht]
    \caption{\label{tab:models}Equilibrium models used for the
      collision (model A) or that are discussed in figure~\ref{fig:2}
      (models B and C). Listed are: the value of the gravitational
      (ADM) mass, the total rest-mass, the radius of the star, the
      central rest-mass density and the polytropic constant $K$. For
      each model we also report the maximum allowed ADM and
      rest-masses for a TOV having the same polytropic constant.}
    \begin{indented}
    \item[]
    \begin{tabular}{cccccccc}
    \br
    Model & $M_{\mathrm{ADM}}$ & $M_b$ & $R$ (km) & $\rho_c$ & $K$ &
$M_{\mathrm{ADM,max}}$ & $M_{b,{\mathrm{max}}}$ \\
    \mr
A   & $0.732$ & $0.760$ & $14.761$ & $0.00058$ & $80$ & $1.464$ & $1.609$
\\
B   & $1.505$ & $1.514$ & $9.135$  & $0.00963$ & $145$ & $1.972$ & $2.166$
\\
C   & $1.460$ & $1.547$ & $19.003$ & $0.00055$ & $155$ & $2.038$ & $2.240$
\\
    \br
    \end{tabular}
    \end{indented}
\end{table}

The stars constructed in this way are then boosted towards each other
along the $z$-axis with a velocity $v_0=0.15$, which is similar to the
free-fall velocity as computed from the Newtonian expression for a
point-particle. With these choices the only remaining free parameter
needed to characterize the initial data is the central rest-mass
density of the two NSs, $\rho_c$, which we use as our critical
parameter (We note that the value chosen for the initial velocity does
influence qualitatively the results obtained and indeed it can act as
a critical parameter in a sequence at constant
$\rho_c$~\cite{Jin:07a}).

\subsection{Entropy of the critical solution}

A convenient way to study the properties of the critical solution is
to characterize its thermodynamical properties and in particular its
entropy. The basic idea is that the metastable (critical) solution is
simply an equilibrium solution on the unstable branch of TOV
configurations and thus that by measuring its entropy it is possible
to relate it to the corresponding equilibrium polytropic
model. Following~\cite{Chandreaskahr39a}, we express the second law of
thermodynamics as
\begin{equation}
\label{eq:entropy}
S \,=\, S_{0} \,+\, C_{\rm V} \; \ln(T) \,-\, R \; \ln(\rho),
\end{equation}
where $S_0$ is an integration constant that we set to zero and where
\begin{equation}
\label{eq:specific_heat}
C_{\rm V}\,=\, \left(\frac{d\epsilon}{dT}\right)_{\rm v}
\end{equation}
is the specific heat capacity at constant pressure. and $R$ the gas
constant. In the case of an ideal gas $R$ is related to the specific
heat capacities through the relation
\begin{equation}
\label{eq:gas_constant}
R \,=\, C_{\rm p}\,-\,C_{\rm V} \,=\, \left(\frac{dh}{dT}\right)_{\rm p} -
\left(\frac{d\epsilon}{dT}\right)_{\rm v},
\end{equation}
where $h = 1 + \epsilon + p/\rho$ is the specific enthalpy. Recalling
that $\epsilon = C_{\rm V} T$ and $\Gamma = C_{\rm p}/C_{\rm V}$, the second
law of thermodynamics (\ref{eq:entropy}) for a polytropic EOS is
simply given by
\begin{equation}
\label{eq:entropy_2}
S \,=\, C_{\rm V} \; \ln \left(\frac{K}{C_{\rm V}(\Gamma - 1)}\right)\,,
\end{equation}
or equivalently
\begin{equation}
\label{eq:kappa_1}
K \,=\, R \; \exp\left(\frac{S}{C_{\rm V}}\right).
\end{equation}
Stated differently, the polytropic constant reflects all of the
changes in the entropy of the system, so that
expression~(\ref{eq:kappa_1}) allows a simple connection between the
entropy of the critical solution, which we measure as proportional to
$K = p / \rho^\Gamma$, and the properties of a corresponding
equilibrium TOV model.

In practice the polytropic constant can change enormously across the
star especially after the collision and since we are interested only
in global quantities we use a volume-averaged polytropic constant
\begin{equation}
\label{eq:kappa_avg}
\langle K\rangle \,=\, \frac{\int_\Omega p / \rho^{\Gamma} dV}{\int_\Omega dV},
\end{equation}
and perform the volume integration not across the whole star but over
a volume $\Omega$ where the rest-mass density is larger than $10\%$ of
the initial central one. This choice removes the difficulties with
possible divergences near the stellar surface and we have verified
that is robust against different values of the threshold density.

\section{Results}
\label{sec:results}

We next discuss the dynamics of the collision and the properties of
the critical solution as they appear from our numerical simulations.
\subsection{Dynamics of the collision}

The basic dynamics of the process is rather simple to describe. As the
two NSs are accelerated towards each other by the initial boost and
the mutual gravitational attraction, they collide, leading to a merged
object which is wildly oscillating and with a mass which is above the
maximum mass of the initial configuration. Depending on whether the
initial central density is larger or smaller than the critical one,
the metastable solution either collapses to a black hole (supercritical
solutions) or it expands to a new stable stellar solution (subcritical
solutions).

Before entering the details of the discussion it may be useful to
remark that the simulation of the head-on collision of two NSs in the
neighborhood of the critical solution is a very demanding calculation,
even for modern general-relativistic hydrodynamical codes. This is
because adaptive mesh refinements do not provide any significant speed
up and, at the same time, rather high level of resolutions are needed
to capture the dynamics faithfully. In spite of these computational
difficulties mentioned above, using the \texttt{Whisky2D} code we are
able to identify the critical value for $\rho_c$ with an accuracy of
$11$ significant digits, a level of precision never achieved before in
the study of NS head-on collisions. More specifically, we have
measured the critical central density
\begin{equation}\label{eq:rho.critical}
\rho_c^\star = ( 5.790998966725 \pm 0.00000000003 ) \times 10^{-4}.
\end{equation}
as the midpoint between the largest central rest-mass density among
the subcritical models and the smallest central rest-mass density
among the supercritical ones. In other words, binaries with
$\rho_c>\rho_c^\star$ have been simulated to collapse to a black hole,
while binaries with $\rho_c<\rho_c^\star$ have been computed to expand
to a stable star.

As pointed out by~\cite{Jin:07a}, this value will ultimately depend on
the numerical resolution used and the other discretization
parameters. Nevertheless, given a set of initial data and numerical
setup, (\ref{eq:rho.critical}) gives a precise measure of how close we
are able to get to the critical solution.

\begin{figure}[ht]
  \begin{center}
  \includegraphics[width=9.0cm]{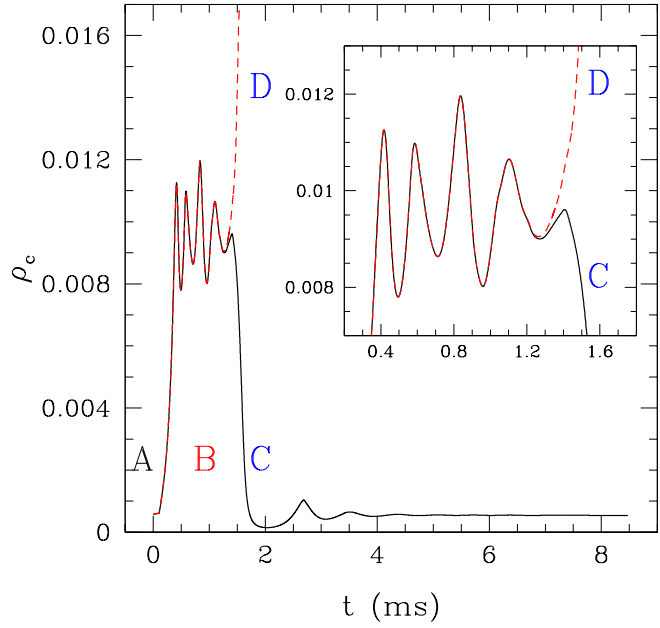}
\end{center}
  \caption{\label{fig:0} Evolution of the maximum rest-mass density of
    the most massive subcritical model with initial value $\rho_{\rm
      c} = 0.000579099896670$ (black solid line) the least massive
    supercritical model with initial value $\rho_{\rm c} =
    0.000579099896675$ (red dashed line). Also highlighted are the
    four different phases of the dynamics. The first one corresponds
    to the initial configuration of the system (labelled as
    ``A''). After the collision, a new metastable solution is created
    during which the central density exhibits violent oscillations
    (labelled as ``B''); the subrcritical and supercritical solutions
    are essentially indistinguishable during this stage. Finally the
    subcritical and supercritical solutions separate, with the first
    one relaxing to a stable expanded configuration (labelled as
    ``C''), while the second collapses to a black hole (labelled as
    ``D'').}
\end{figure}

Compressed in figure~\ref{fig:0} is a considerable amount of
information about the criticality of the head-on collision. More
specifically, we show the time evolution of the central rest-mass
density of the heaviest subcritical model (black solid line) and
lightest supercritical model (red dashed line) computed. Overall, we
can distinguish three different phases of the dynamics. In the first
one (marked with ``A'' in the figure) the central density increases
from its initial value to a maximum one reached, which is attained
when the stellar cores merge.

\begin{figure*}
 \begin{center}
 \includegraphics[width=0.495\textwidth]{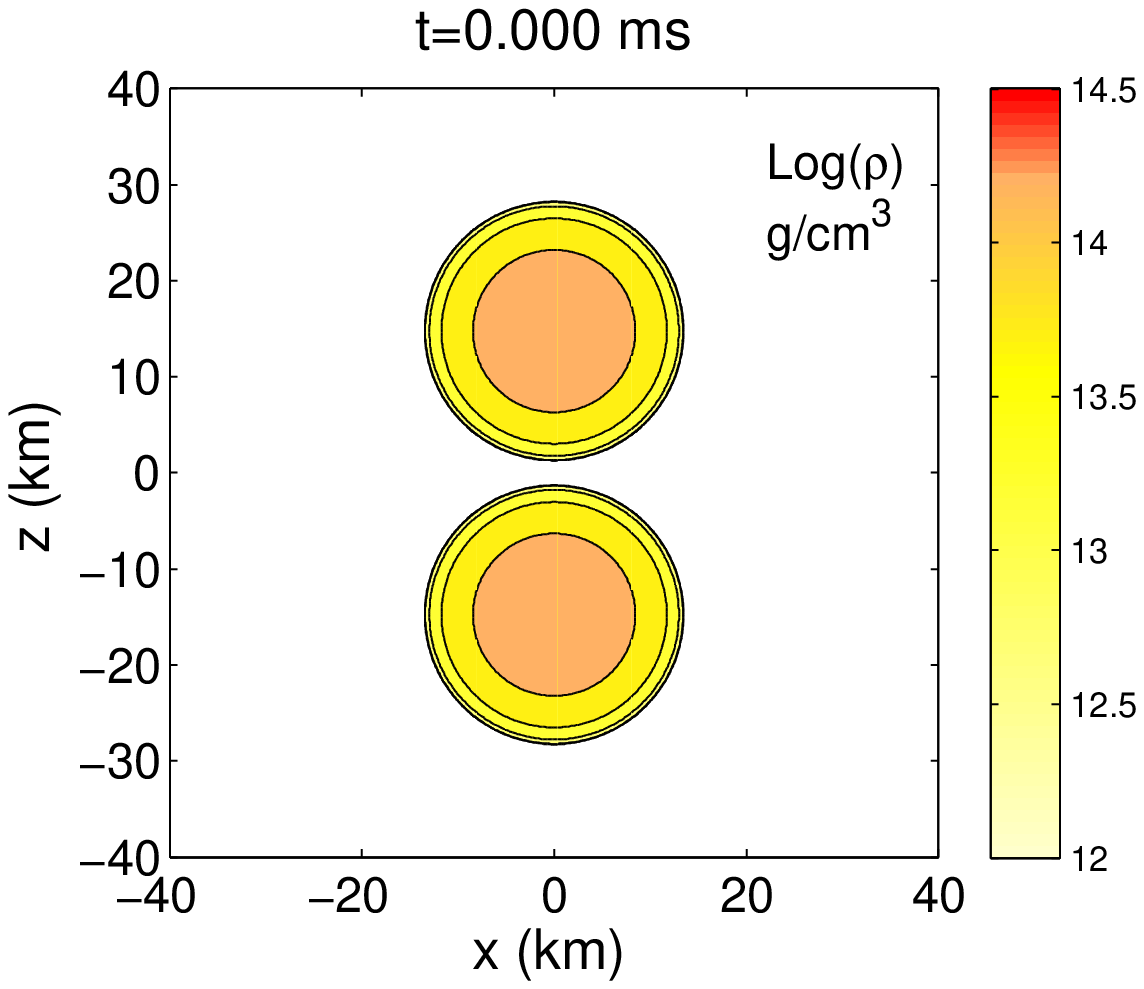}
 \includegraphics[width=0.495\textwidth]{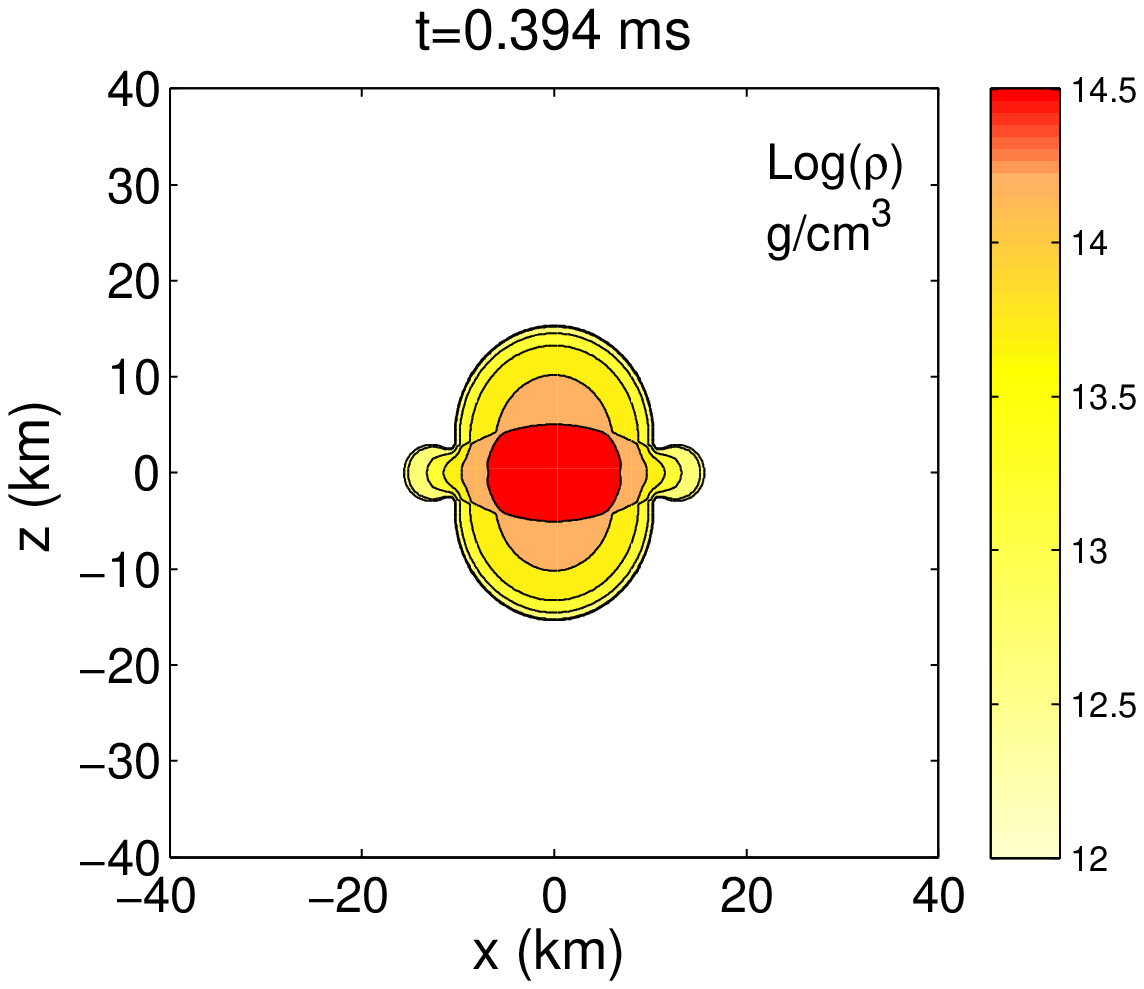}
 \includegraphics[width=0.495\textwidth]{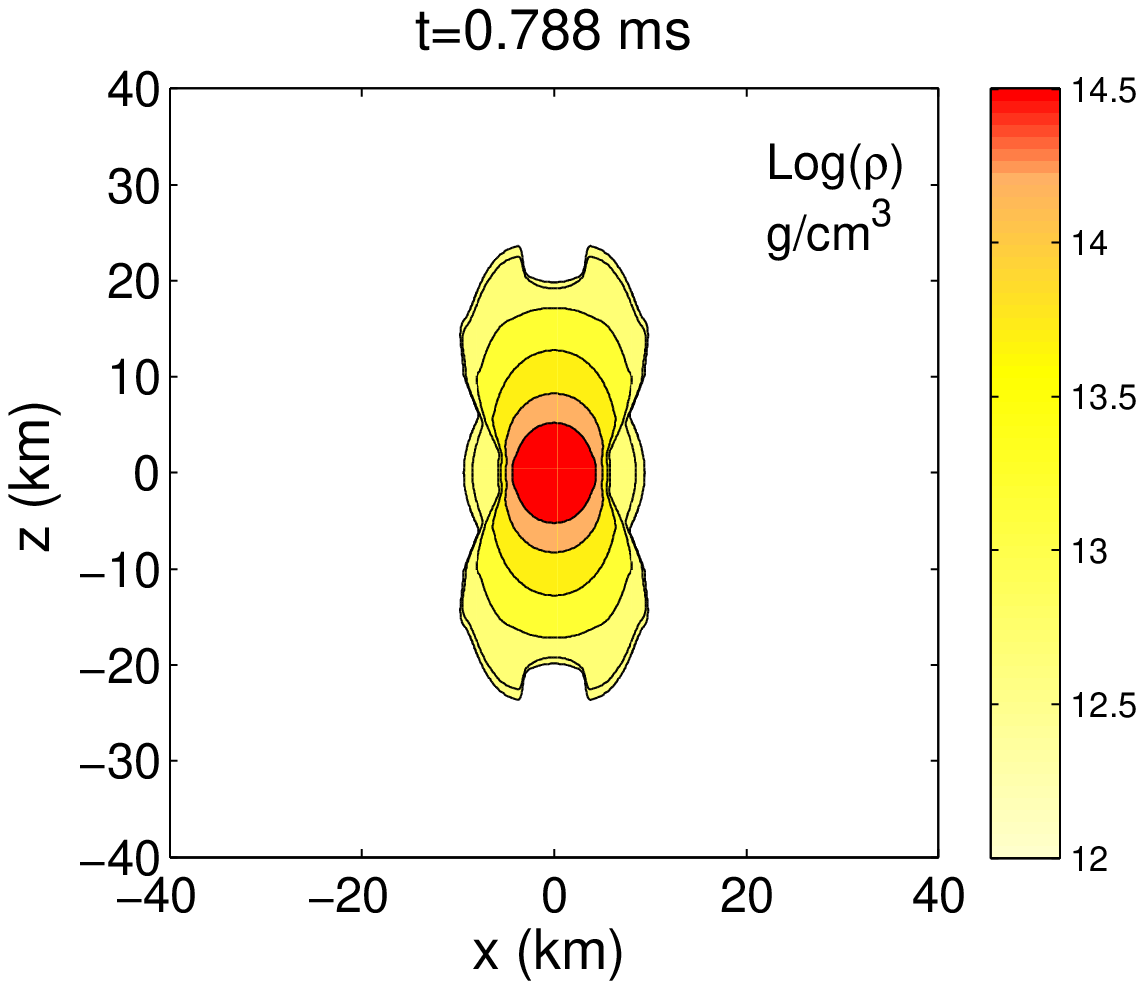}
 \includegraphics[width=0.495\textwidth]{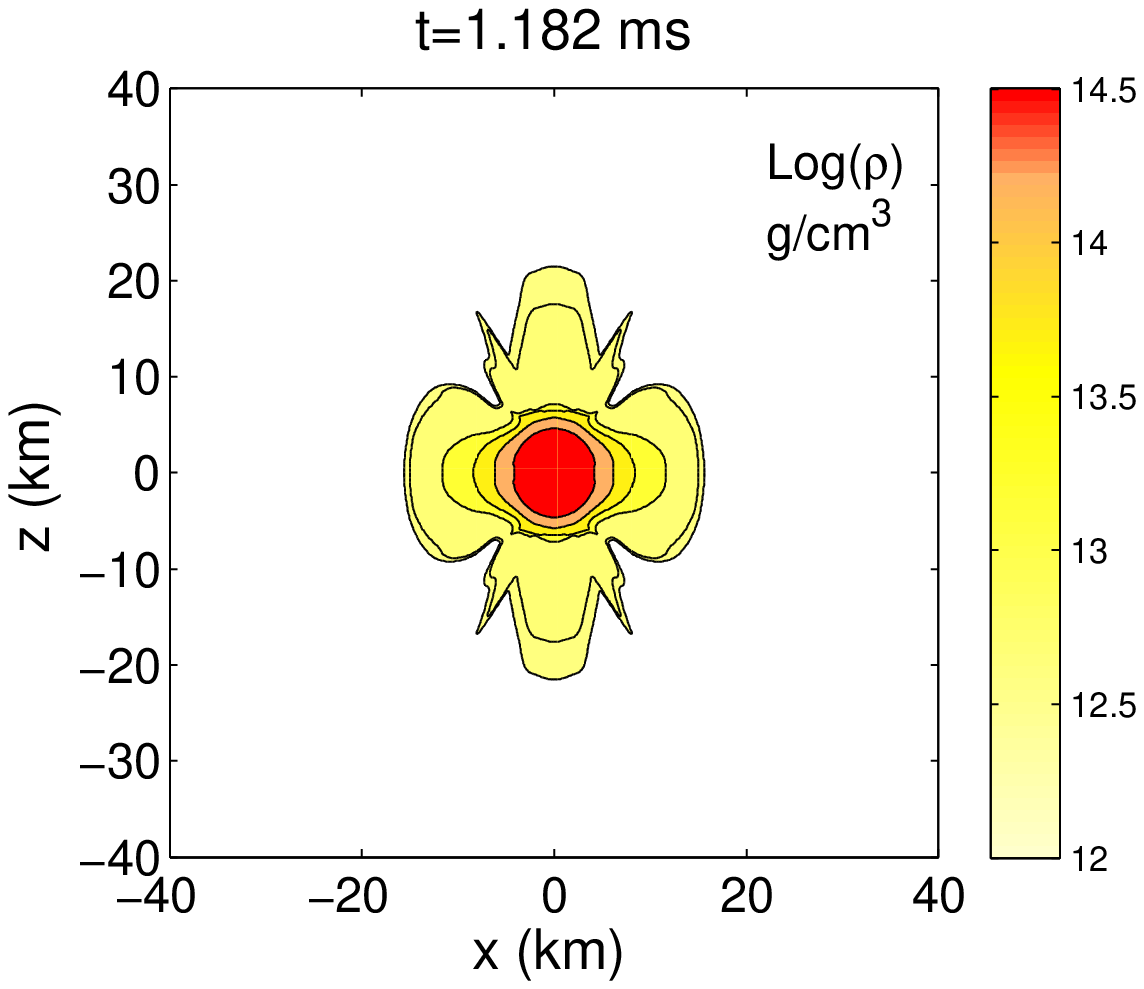}
 \includegraphics[width=0.495\textwidth]{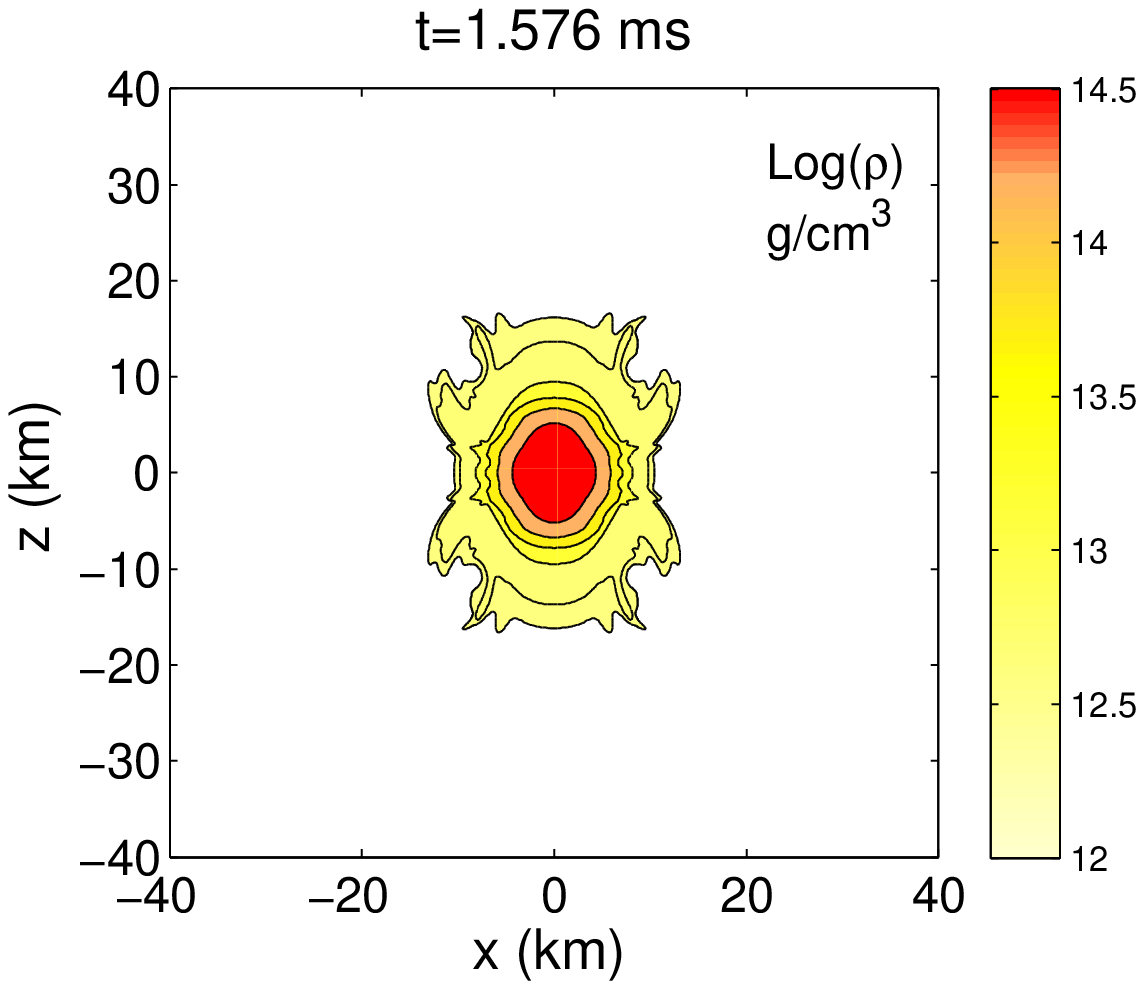}
 \includegraphics[width=0.495\textwidth]{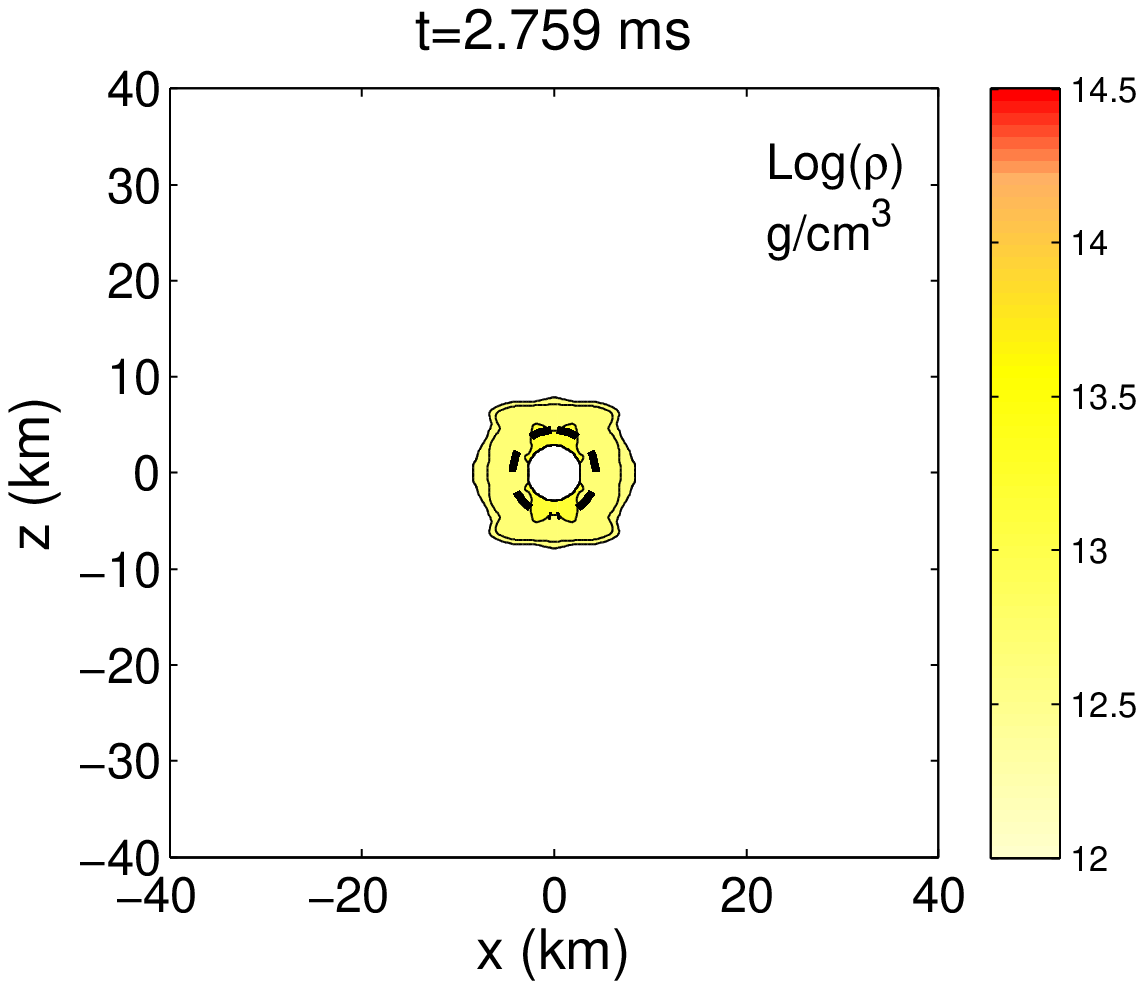}
\end{center}
\vskip -0.25cm
\caption{\label{fig:contplot1} Isodensity contours in the $(x,z)$
  plane of the least massive supercritical solution ($\rho_c =
  0.000579099896675$). The corresponding times are shown at of each
  panel, while the color-code for the rest-mass density is indicated
  to the right. The isolines are shown for the values of $\rho =
  10^{12.6}, 10^{12.7}, 10^{13.2}, 10^{13.7}, 10^{14.2}$ and
  $10^{14.7}$ $\mathrm{g}/\mathrm{cm}^3$. The second frame ($t =
  0.394$ ms) is taken during the merge process. The next five frames
  illustrate the star during the metastable equilibrium. Finally the
  last frame ($t = 2.759$ ms) shows the solution after the formation
  of a black hole, whose apparent horizon is shown with a thick black
  dashed line.  }
\end{figure*}
\begin{figure*}
\begin{center}
 \includegraphics[width=0.495\textwidth]{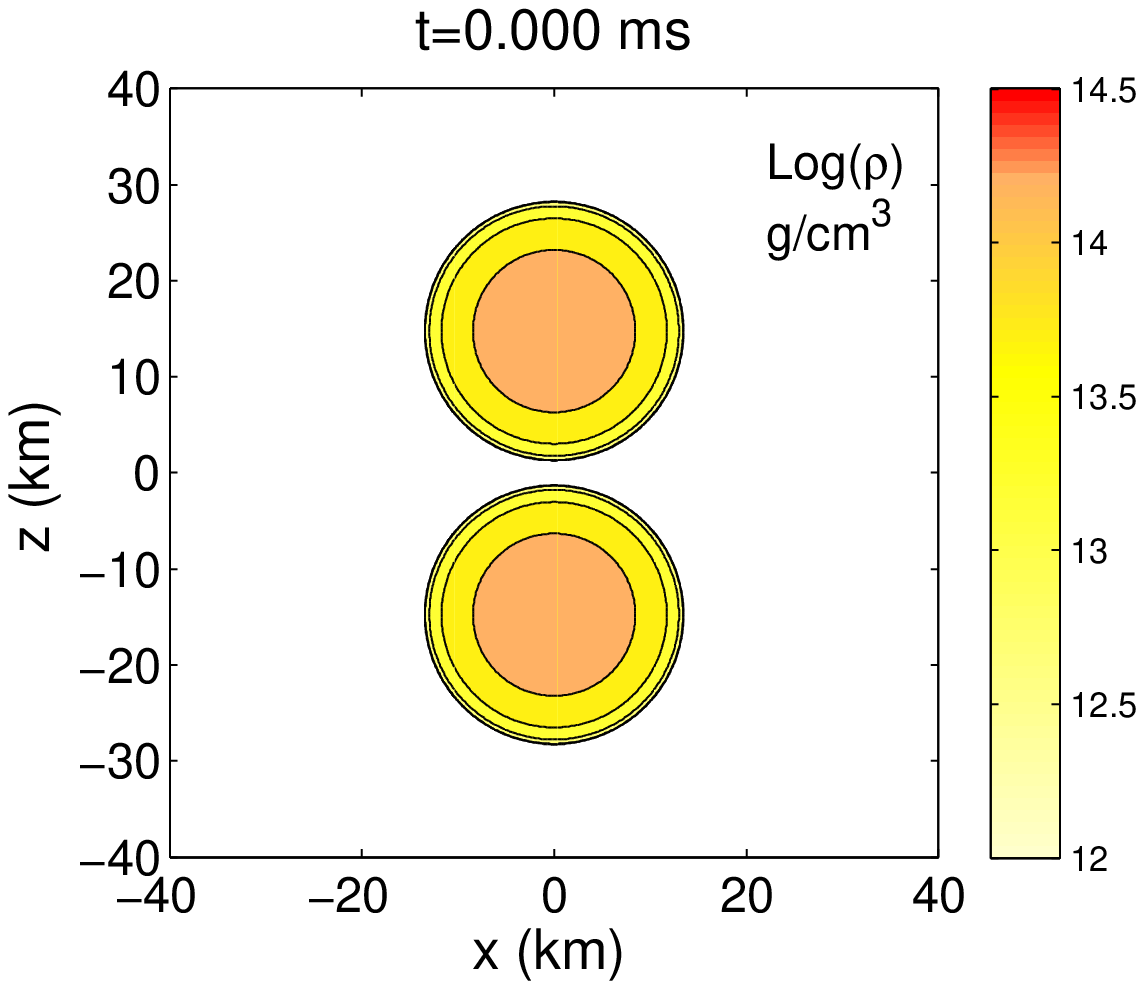}
 \includegraphics[width=0.495\textwidth]{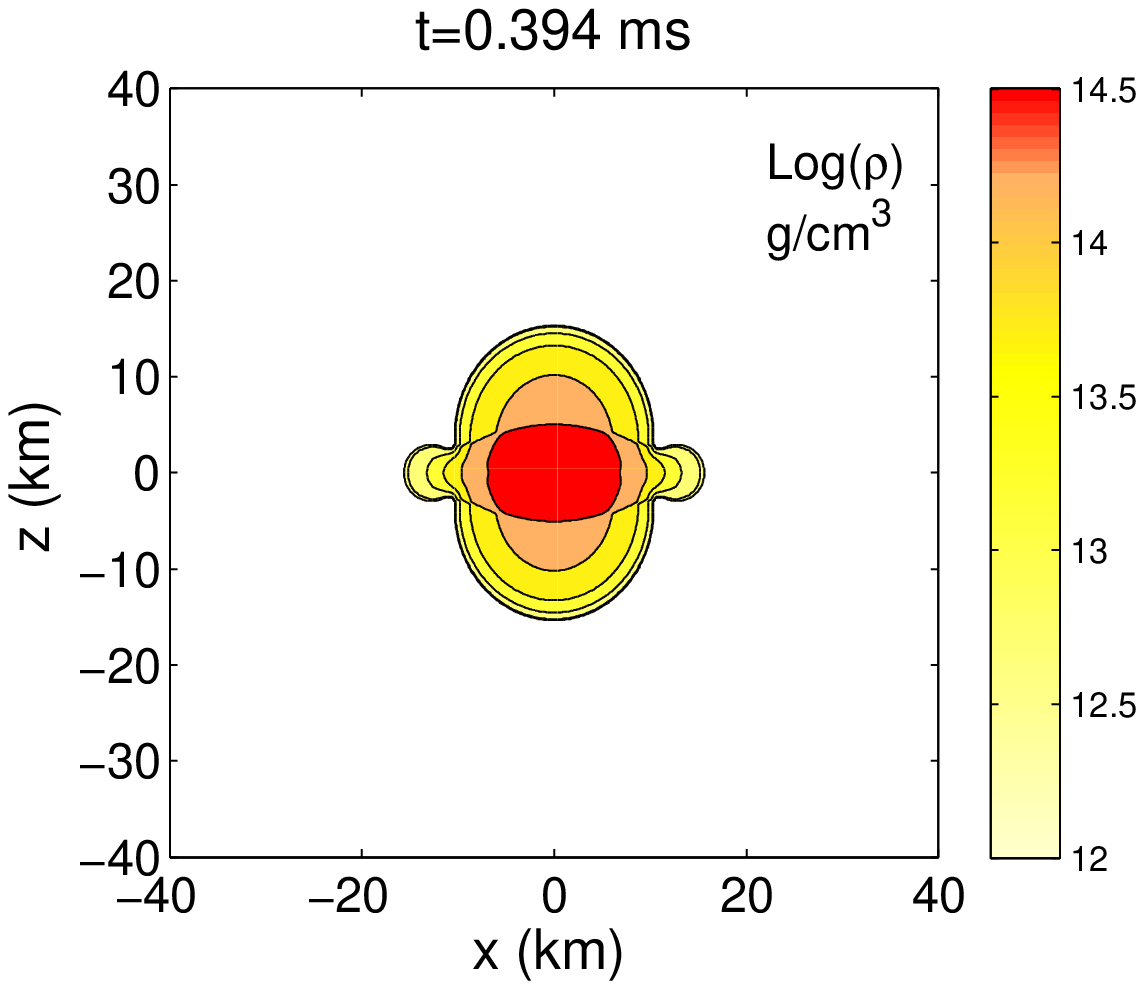}
 \includegraphics[width=0.495\textwidth]{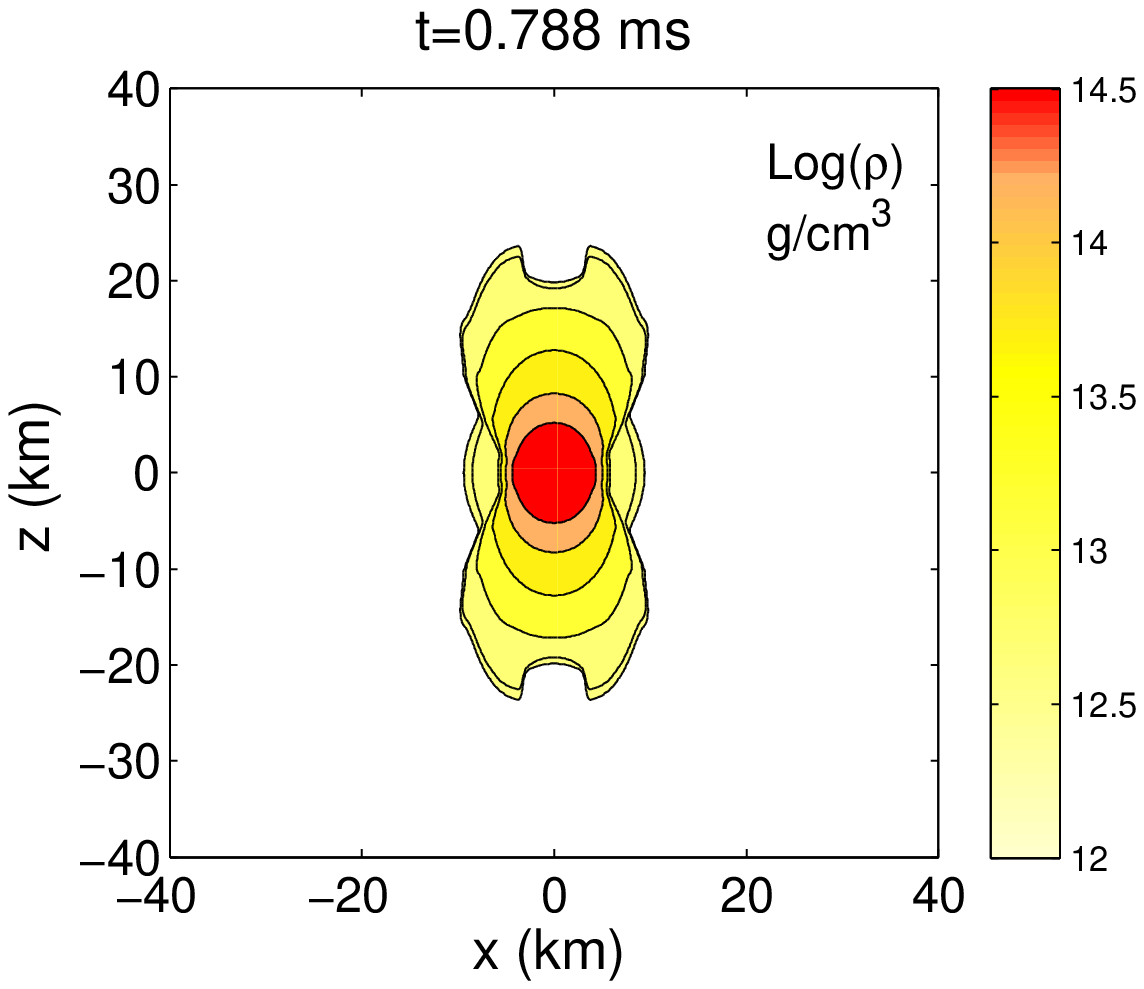}
 \includegraphics[width=0.495\textwidth]{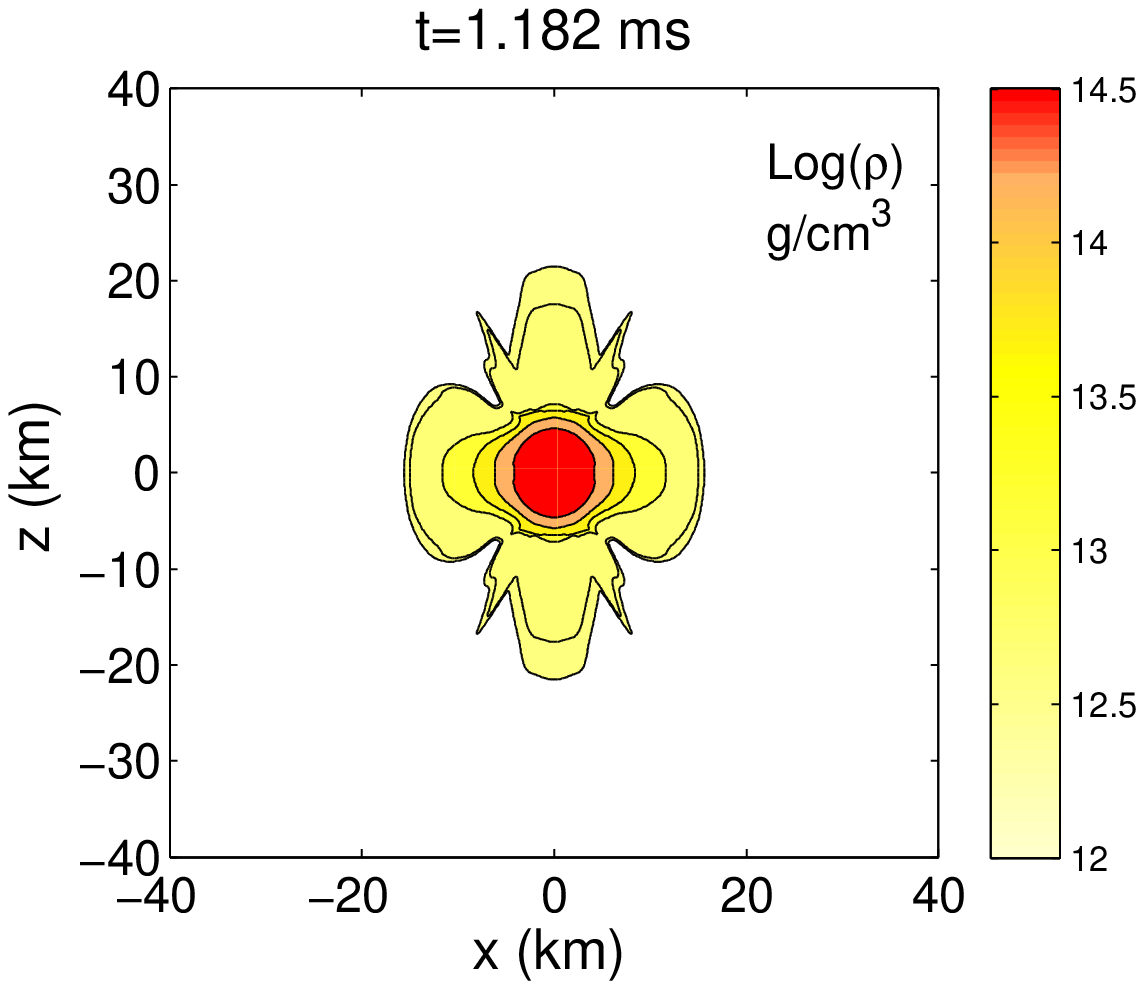}
 \includegraphics[width=0.495\textwidth]{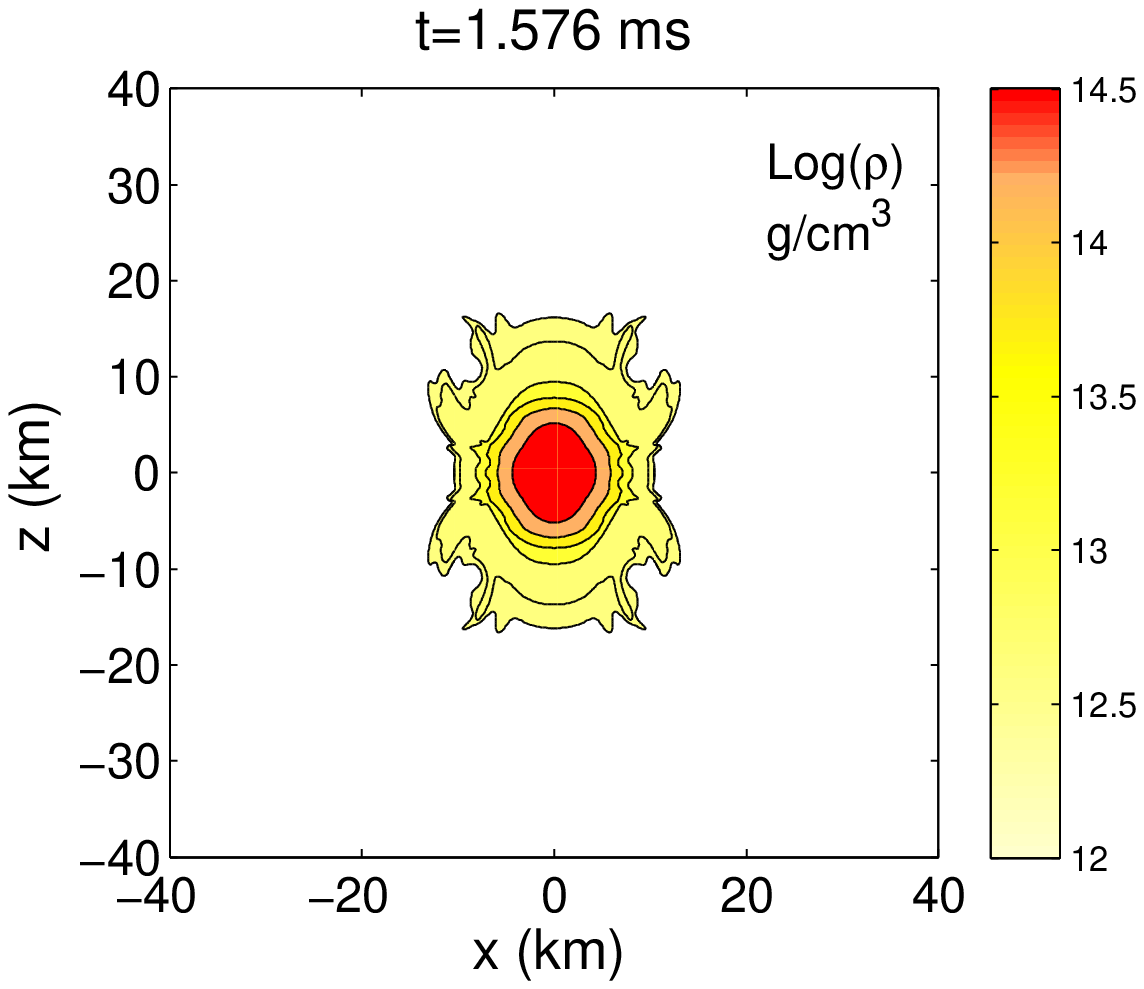}
 \includegraphics[width=0.495\textwidth]{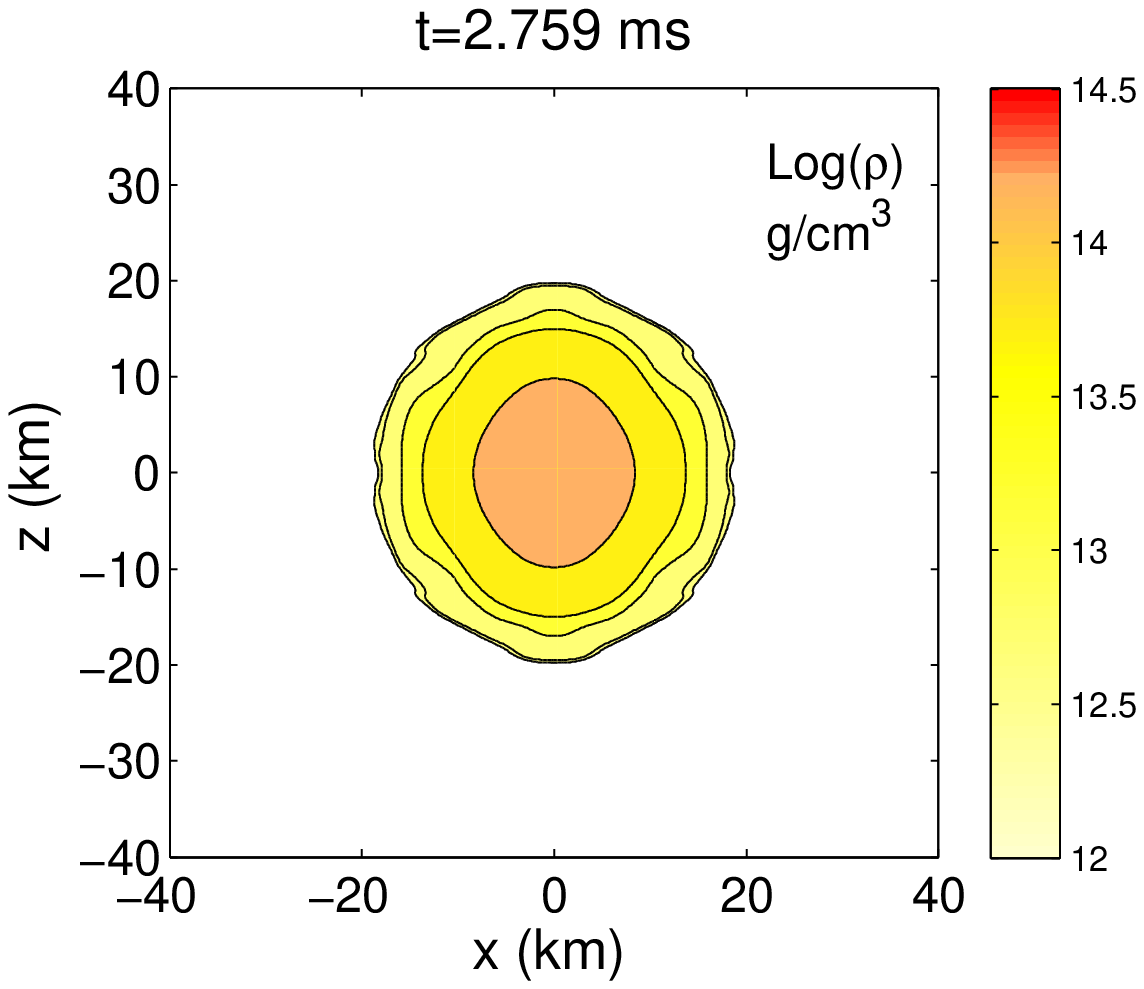}
\end{center}
\vskip -0.25cm
\caption{\label{fig:contplot2} The same as figure~\ref{fig:contplot1},
  but for the most massive subcritical model ($\rho_c =
  0.000579099896670$). The first five frames are similar (although not
  identical) to the corresponding ones in figure~\ref{fig:contplot1}
  since they refer to the metastable evolution when the two solutions
  are essentially indistinguishable. The last frame ($t\,=\,2.759$ ms)
  shows a new NS produced from the migration of the metastable object
  to the stable branch of the equilibrium solutions.}
\end{figure*}

The second phase (marked with ``B'') starts from $t\sim 0.4$ ms and is
characterized by strong oscillations around the metastable
equilibrium. During these first two phases the subcritical solution d
the supercritical one are essentially indistinguishable, but at the
end of the second phase, \ie at $t\sim 1.2$ ms, the different nature
of the two solutions emerges and the evolutions of the rest-mass
density differ. More specifically, during the third phase (marked
with ``C'' and ``D''), the supercritical solution shows an exponential
increase of the central rest-mass density as a result of the collapse
to a black hole. The subcritical solution, on the other hand, shows a
violent expansion and the central density settles to a value which is
about one tenth of the maximum one attained during the second
phase. Clearly, the most interesting part is obviously the one
corresponding to the second phase, during which the merged object is a
metastable solution in which the central density has strong,
non-harmonic oscillations (see inset of figure \ref{fig:0}).

In order to better describe the dynamics of the system, we show in
figures~\ref{fig:contplot1} and~\ref{fig:contplot2} the evolution of
colour-coded countours of the rest-mass density of the supercritical
and subcritical solutions, respectively, when shown at representative
times in the $(x,z)$ plane.  The first row of panels in figure
\ref{fig:contplot1} shows the initial configuration of the system at
time $t=0$ and a subsequent stage, at time $t = 0.394\ \mathrm{ms}$,
corresponding to when the first maximum in the rest-mass density is
reached (\cf~figure \ref{fig:0}). This time also represents the one at
which the two stellar cores enter in contact and thus marks the
beginning of the metastable equilibrium. During this stage, two strong
shock waves propagate along the $z$-direction, ejecting part of the
matter as shown in the third panel at time $t =
0.788\ \mathrm{ms}$. Most of this matter is still gravitationally
bound and falls back onto the central object creating a new shock wave
(\cf~fourth panel at $t = 1.182\ \mathrm{ms}$). This process is then
repeated multiple times and results in a sequence of bounces until the
object finally collapses to a black hole, as shown in the last panel
at time $t=2.759\ \mathrm{ms}$ and which marks the fate of the
supercritical solution. 

Similarly, figure~\ref{fig:contplot2}, reports representative stages
of the evolution of the subcritical solution. A rapid inspection of
the first five panels of figure \ref{fig:contplot1} indicates they are
very similar to the corresponding ones in
figure~\ref{fig:contplot1}. Indeed, the supercritical and subcritical
solutions are the same to the precision at which we measure the
critical solution [\cf~(\ref{eq:rho.critical})] and we have reported
the panels here exactly to remark the similarity during the first two
stages of the evolution. However, being it a subcritical solution, the
metastable evolution does not end with a black hole formation but,
rather, with a new stable stellar solution. This is shown in the sixth
panel of figure~\ref{fig:contplot2} and refers to a time $t =
2.759\ \mathrm{ms}$, after the metastable star has expanded violently
and when it has reached a new quasi-spherical configuration.

With the exception that we are able to get closer to the critical
solution, much of what reported here confirms what found by Jin and
Suen in~\cite{Jin:07a}. In the following section, however, we discuss
how and why our conclusions about the properties of the critical
solutions differ from those discussed in~\cite{Jin:07a} and
subsequently in~\cite{wan_2008_das}.

\subsection{Nearly-critical solutions}

\begin{figure*}[h]
  \begin{center}
  \includegraphics[width=6.25cm]{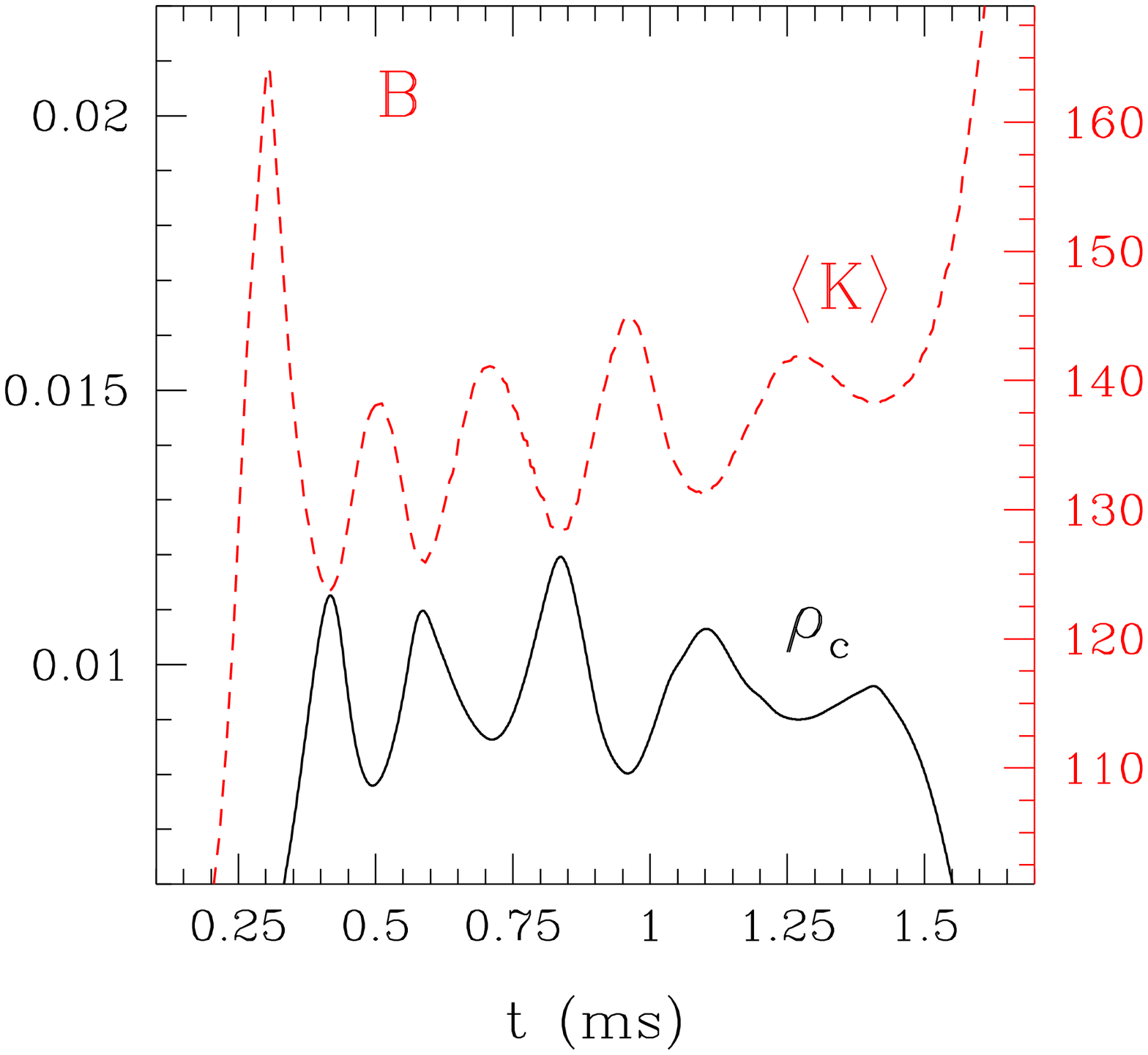}
  \includegraphics[width=6.25cm]{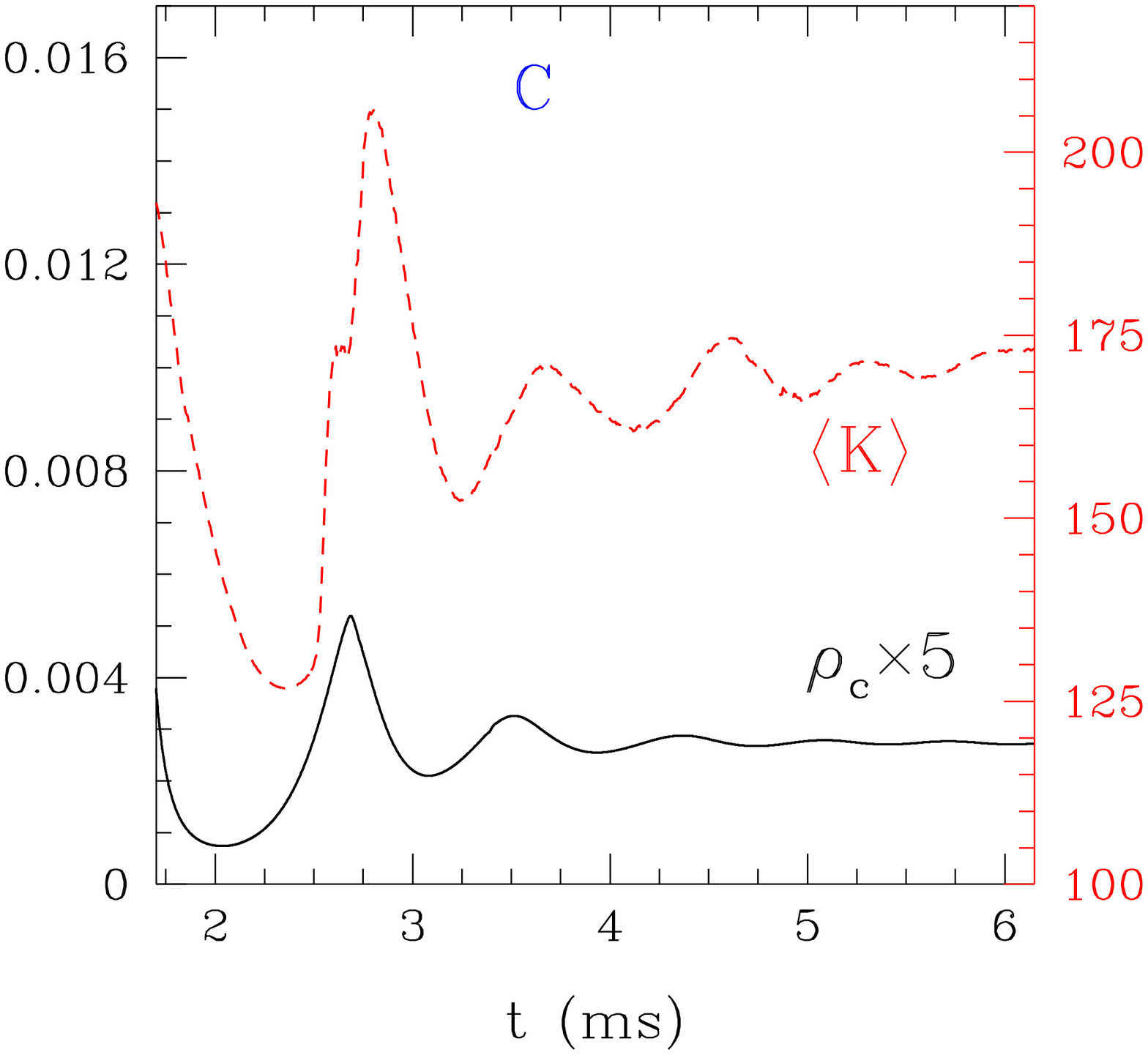}
\end{center}
  \vskip -0.5cm
  \caption{\label{fig:1} \textit{Left panel:} Evolution of the central
    rest-mass density and of the effective polytropic constant
    $\langle K \rangle$ for the most massive subcritical model during
    the metastable equilibrium phase of the dynamics. \textit{Right
      panel:} The same as in the left panel but during the relaxation
    to a stable configuration.}
\end{figure*}

The scope of this section is to show that, contrary to what suggested
in ~\cite{Jin:07a,wan_2008_das}, the metastable object can be
interpreted rather simply as the perturbation of a new equilibrium
configurations of linearly unstable spherical stars. To provide
evidence that this is the case, we have computed the evolution of the
average entropy of the system for a subcritical solution as computed
via the volume-averaged polytropic constant~(\ref{eq:kappa_avg}). This
is shown in figure \ref{fig:1}, where we report both $\langle K
\rangle$ and the central density $\rho_c$ during the metastable
equilibrium or stage ``B'' (left panel) and when the metastable
solution has expanded to recover a stable solution or stage ``3''
(right panel).

Clearly, the two panels show that two quantities are correlated and
indeed in phase opposition -- entropy increases when the density
decreases and viceversa -- as one would expect from the first law of
thermodynamics
\begin{equation}
	T dS = p dV + d Q = p dV\,,
\end{equation}
where the second equality comes from assuming adiabatic transformations.

Using the results shown in figure \ref{fig:1}, it is then possible to
compute a time-averaged value of $\langle K \rangle$ and of the
central density, \ie $\overline{\langle K \rangle}$,
$\overline{\rho_c}$, and thus equilibrium polytropic models with such
polytropic constant and central density. These equilibrium models can
be constructed either relative to the metastable stage or relative to
the final stable stage of the subcritical solutions.

\begin{figure}[ht]
  \begin{center}
  \includegraphics[width=9.0cm]{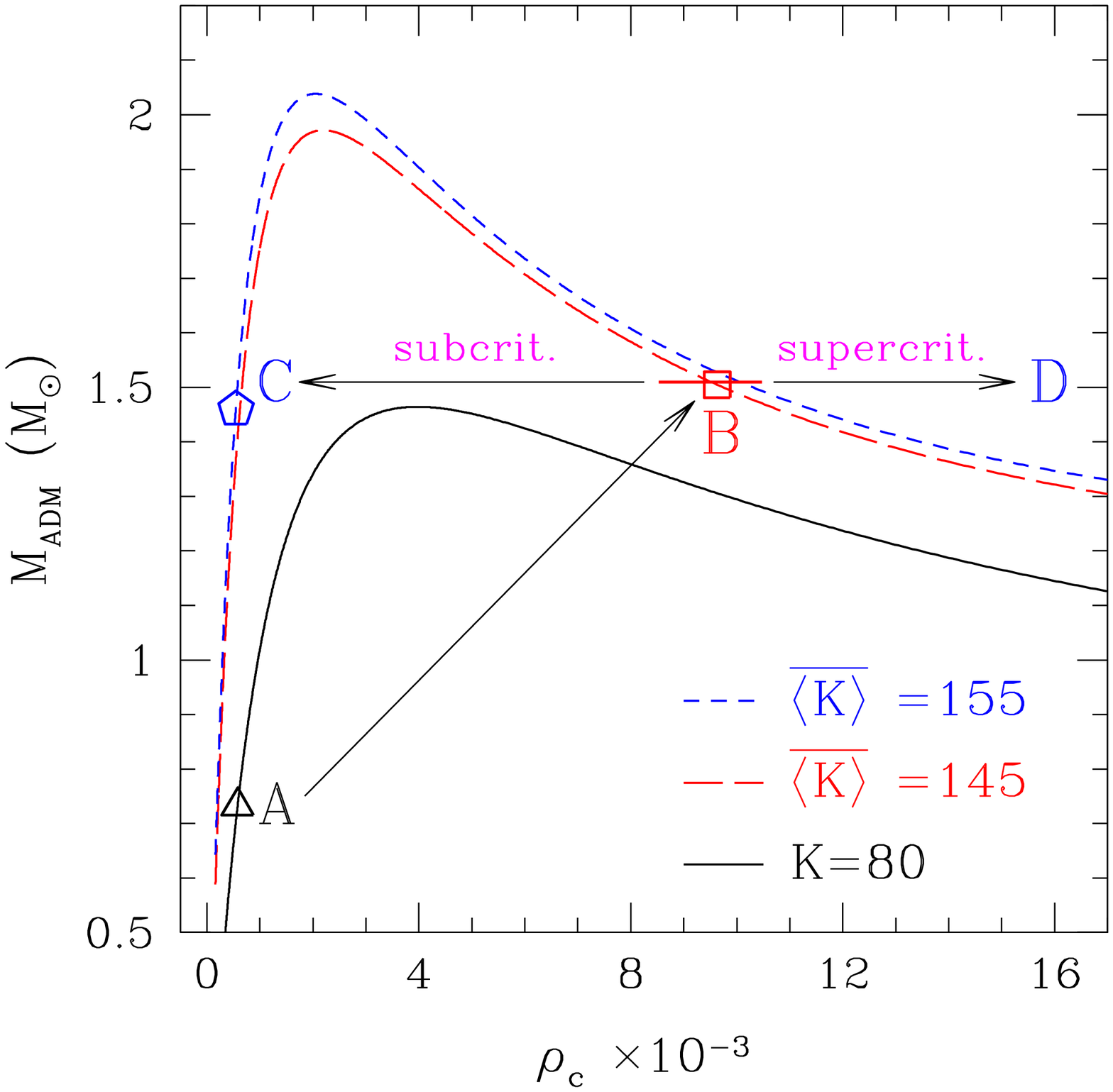}
\end{center}
  \vskip -0.5cm
  \caption{\label{fig:2} Equilibrium sequences of TOV stars with fixed
    polytropic constant in a standard $(\rho_c, M_{\rm ADM})$
    plane. The black solid line refers to a sequence with $K=80$ and
    the black triangle marks the initial critical solution (\ie model
    ``A'' in table~\ref{tab:models}). The red long-dashed line to a
    sequence with $\overline{\langle K \rangle}=145$ and the red
    square shows the equilibrium model having as central density the
    time-averaged central density of the metastable solution (\ie
    model ``B'' in table~\ref{tab:models}). Finally, the blue dashed
    line refers to a sequence with $\overline{\langle K \rangle}=155$
    and the blue pentagon shows the equilibrium model having as
    central density the time-averaged central density of the stable
    solution (\ie model ``C'' in table~\ref{tab:models}). All models
    ``A, B, C'' have the same rest-mass to a precision of $0.4\%$ and
    the arrows show how the collision corresponds to a number of
    transition in the space of configurations.}
\end{figure}

The results of this procedure are summarized in figure~\ref{fig:2},
which reports the equilibrium sequences of TOV stars with fixed
polytropic constant in a standard $(\rho_c, M_{\rm ADM})$ plane. In
particular, the black solid line refers to a sequence with $K=80$ and
the black triangle therefore marks the initial critical solution (\ie
model ``A'' in table~\ref{tab:models}). Similarly, the red long-dashed
line refers to a sequence with a polytropic constant
$\overline{\langle K \rangle}=145$, which therefore coincides with the
time averaged value of $K$ during the metastable state and as deduced
from figure~\ref{fig:1}. The red square shows therefore the
equilibrium model having as central density the time-averaged central
density of the metastable solution (\ie model ``B'' in
table~\ref{tab:models}). Finally, the blue dashed line refers to a
sequence with a polytropic constant $\overline{\langle K
  \rangle}=155$, which therefore coincides with the time averaged
value of $K$ during the stable stage of the subcritical
solutions. (\cf figure~\ref{fig:1}). The blue pentagon shows therefore
the equilibrium model having as central density the time-averaged
central density of the stable solution (\ie model ``C'' in
table~\ref{tab:models}). It is important to remark that because models
``A, B, C'' are determined after fixing the polytropic constant and
the central rest-mass density, they are not guaranteed to have the
same total baryon mass. In practice, however, they do have the same
rest-mass with a precision of $0.4\%$ (of course $2 M_{b, A} \simeq
M_{b,B} \simeq M_{b,C}$). This is not a coincidence but a clear
evidence of the common link among the three models.

What is eloquently shown in figure~\ref{fig:2} can also be stated
summarized as follows. The head-on collision of two NSs near the
critical threshold can be seen as series of transitions in the space
of configurations from an initial stable model ``A'' over to a
metastable model ``B'' which has the same rest-mass but larger
gravitational mass as a result of the conversion of the kinetic energy
into internal energy via large shocks. Because model ``B'' is on the
linearly unstable branch of the equilibrium configurations, it can
exhibit a critical behaviour (this was shown in great detail in paper
I) and thus subcritical solutions will expand and move the stable
branch of equilibrium solutions (model ``C''), while supercritical
solutions will collapse to a black hole (solution ``D'').

In the light of this interpretation, the conclusion drawn by Jin and
Suen~\cite{Jin:07a} that the merged object is far from being a TOV
because it promptly collapses even though its total rest-mass is
smaller than the corresponding maximum mass, does not appear to be the
correct one. Indeed, the transition highlighted in figure~\ref{fig:2}
clearly shows that even a sub-massive TOV can be brought over the
stability threshold to collapse to black hole by simply increasing its
gravitational mass, namely by increasing its internal energy.

Additional evidence that the merged object is indeed a perturbed TOV
comes from analyzing the oscillation frequencies measured over the
metastable stage. Despite the latter is rather short and the
eigenfrequencies are consequently not very accurate, they agree well
with the ones obtained from the linear perturbation theory for the
corresponding TOV model.  This is reported in table
\ref{tab:frequencies}, which collects the oscillation frequencies as
computed from a Fourier analysis of the central rest-mass density of
the largest subcritical solution. Because these frequencies with their
error-bars are within the expected ones, we cannot confirm the claim
made in~\cite{wan_2008_das} that the frequencies of the critical
solution are 1 or 2 orders of magnitude smaller than the corresponding
equilibrium ones. Rather, we conclude that the metastable critical
solution is indeed only a perturbed TOV star.

\begin{table}
\caption{\label{tab:frequencies} Eigenfrequencies of the modes of the
  critical solution during the metastable phase of the dynamics as
  computed from the evolution of the the central rest-mass
  density. Also indicated are the first overtones of the fundamental
  mode for a TOV star constructed with $K$ and $\rho_c$ equal to the
  time-averages of $\langle K \rangle$ and $\langle\rho_c\rangle$
  during the metastable phase. Despite the large uncertainty due to
  the very short integration time, the match between the two set of
  eigenfrequencies is very good.}
\begin{indented}
	\item[]
	\begin{tabular}{lll}
	\br
	Mode & Observed freq. (kHz) & TOV freq. (kHz) \\
	\mr
	H1   & $4.8\pm 1.2$ & $3.95$ \\
	H2   & $7.2\pm 1.2$ & $6.86$ \\
	H3   & $9.6\pm 1.2$ & $9.42$ \\
	H4   & $12.0\phantom{.} \pm 1.2$ & $11.85$ \\
	\br
	\end{tabular}
	\end{indented}
\end{table}

\begin{figure*}
  \begin{center}
    \includegraphics[width=6.25cm]{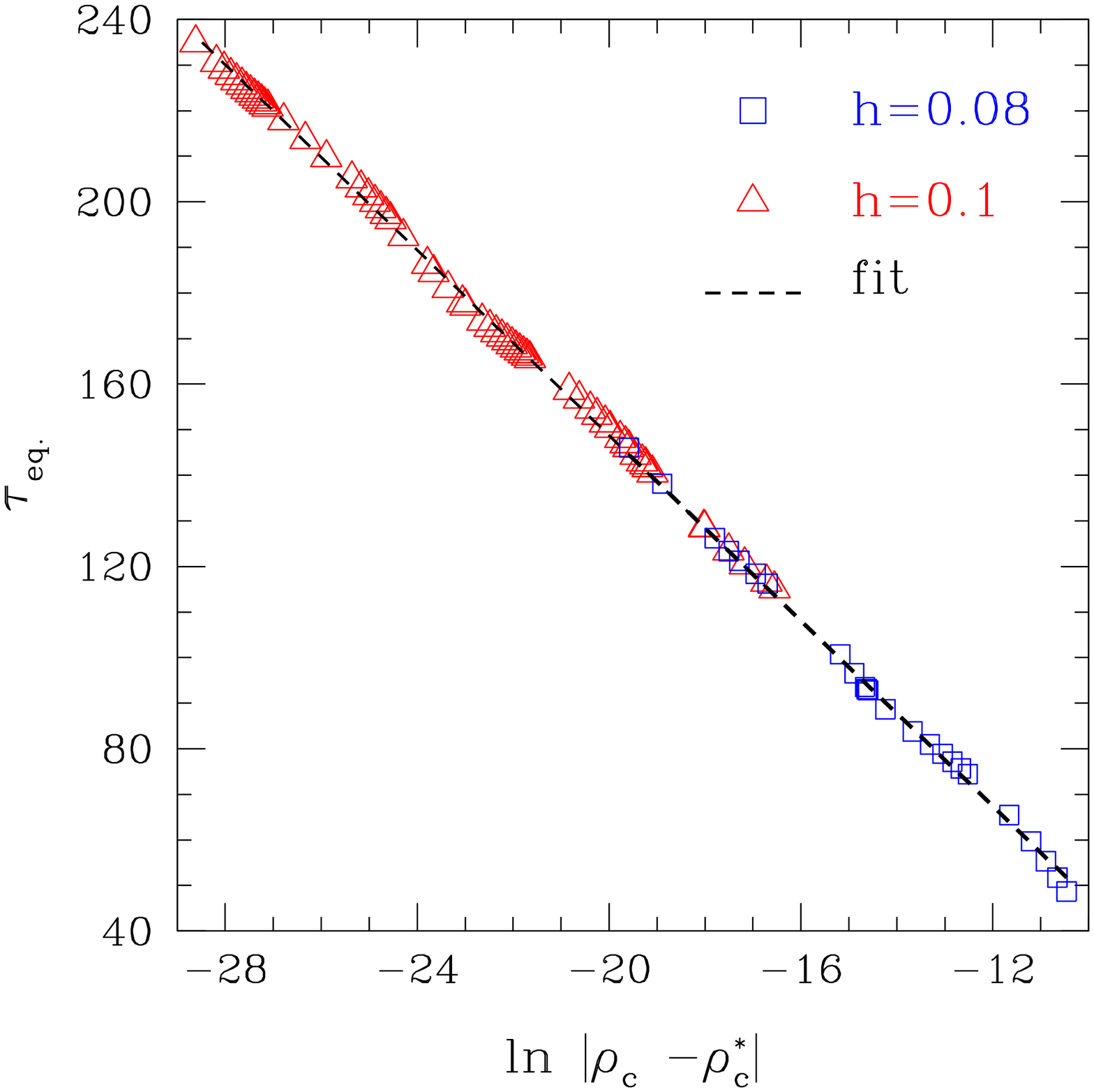}
    \includegraphics[width=6.25cm]{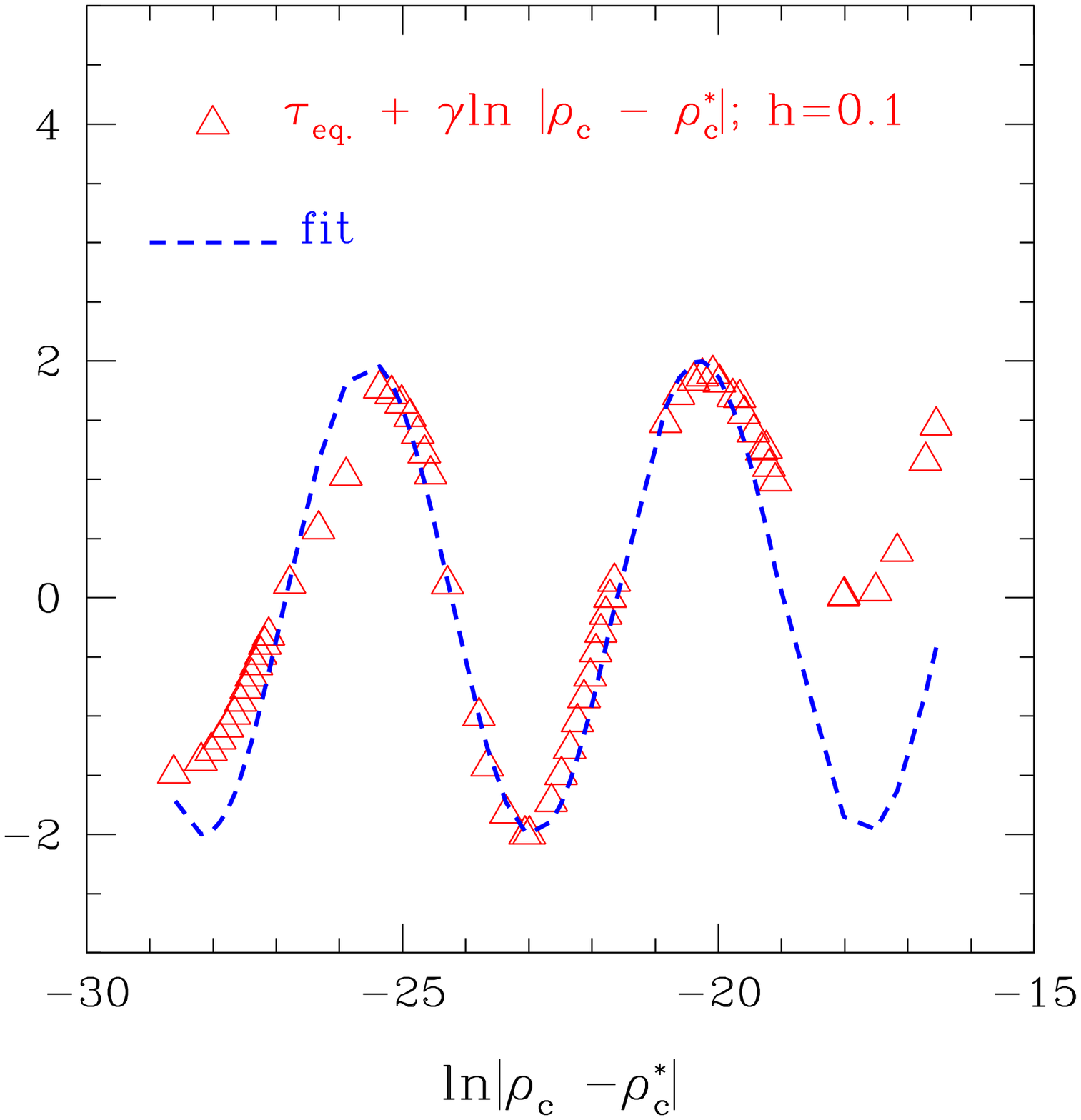}
\end{center}
  \vskip -0.5cm
  \caption{\label{fig:5} \textit{Left panel:} Survival time of the
    metastable solution plotted against the logarithm of the
    difference between the initial central density of the stars and
    the critical one. The red triangles represent the data points
    obtained with a grid resolution of $h=0.1$ and the blue squares
    represent the ones obtained with $h=0.08$. The black dashed line
    represents the expected power-law scaling. \textit{Right panel:}
    Harmonic fluctuations in the critical exponent $\gamma$ as
    obtained after subtracting the power-law scaling from the data
    points computed with $h=0.1$ (red triangles). Indicated with a
    blue dashed line is the fitting sine function.}
\end{figure*}

\subsection{On the critical exponent and its fluctuations}

The theory of critical phenomena predicts a precise scaling relation
between the survival time of the nearly-critical solutions, namely the
time over which a metastable equilibrium exists, and the distance from
the critical solution. At lowest order this scaling relation is a
simple power-law of the type
\begin{equation}
\label{eq:life_time}
\tau_{\rm eq} \simeq - \gamma \ln | \rho_c - \rho^\star_c | + \mathrm{const}\,.
\end{equation}

Following~\cite{Jin:07a}, we measure the survival time by considering
the function $\zeta(t) = (\alpha-\alpha^\star)/\alpha^\star$, where
$\alpha$ is the lapse function at the coordinate origin of a given
simulation and $\alpha^\star$ is the lapse of the best numerical
approximation of the critical solution. We set $\tau_{\mathrm{eq}}$ to
be the first (coordinate) time at which $\zeta(\tau_{\mathrm{eq}})
\geq 0.05$; as discussed in~\cite{Jin:07a}, the determination of the
critical exponent does not depend sensitively on this cut-off
time. Finally, we compute $\gamma$ performing a linear least-square
fit of (\ref{eq:life_time}) on the data points. The results of this
process are shown in the left panel of figure \ref{fig:5}, where we
report with red triangles the values of $\tau_{\mathrm{eq}}$ as
computed from about $60$ simulations having different initial central
density. As we will discuss below, such a large number of data points
is necessary not only to measure accurately the exponent $\gamma$, but
also to determine whether nonlinear corrections to
expression~(\ref{eq:life_time}) should be considered. In this way we
have computed the critical exponent to be $\gamma = 10.004$, which
agrees within $8.4\%$, with the value computed by~\cite{Jin:07a}. As a
further validation, we have computed the critical exponent also for a
(smaller) set of simulations carried out at a higher resolution (\ie
$h=0.08$ vs $h=0.1$) and these are shown as blue squares. These
higher-resolution simulations predict a critical exponent of $10.303$,
thus with a difference of $2.9\%$ from the lower-resolution ones.

A more careful analysis of the data for the survival time reveals that
relation~(\ref{eq:life_time}) is well reproduced by the data, but also
that the latter show additional, fine-structure features which are are
not accounted for in~(\ref{eq:life_time}). In particular, it is
apparent already at a visual inspection that the critical exponent
also shows a periodic change as the solution approaches the critical
one. This is highlighted in the right panel of figure \ref{fig:5},
where we show the deviations of the data from the relation
(\ref{eq:life_time}), and where it is apparent that these deviations
are essentially harmonic in the range in which data is available (for
simplicity and to increase the precision of the fit, we have
considered only the low-resolution data, for which more simulations
are available). As a result, we can correct the scaling
relation~(\ref{eq:life_time}) with a simple expression of the type
\begin{equation}
\label{eq:life_time_n}
\tau_{\rm eq} \simeq - \gamma \ln | \rho_c - \rho^\star_c | + c_1 \sin\big(c_2 \ln |
\rho_c - \rho^\star_c | + c_3\big) + \mathrm{const}\,.
\end{equation}
where $c_1\simeq 2.0$, $c_2\simeq 1.2$ and $c_3\simeq
0.8$. Interestingly, this fine structure of the time-scaling relation
has been observed also in in the critical collapse of scalar
fields~\cite{Gundlach97f,Hod97}, but has never been reported before
for perfect fluids, although it may be present also in the data
of~\cite{Jin:07a} (\cf~their figure 4).  Because the scaling relation
(\ref{eq:life_time}) is derived after performing a linear analysis
near the critical solution~\cite{Hara96a}, the additional oscillation
captured in expression~(\ref{eq:life_time_n}) is a purely nonlinear
effect which has not been yet fully explained.

\section{Conclusions}\label{sec:conclusions}
Critical phenomena are ubiquitous in many different branches of
physics and are of great interest in general relativity where they are
associated with phase transition of families of solutions.  With the
goal of studying the occurrence of type-I critical collapse, we
considered the head-on collision of equal mass, nonrotating NSs
boosted towards each other. After fixing the initial velocity of the
stars, we evolved numerically a great number of configurations with
different initial central rest-mass density using the 2D general
relativistic hydrodynamical code \texttt{Whisky2D}.

Overall, the basic dynamics of the process is rather simple: As the
two NSs are accelerate towards each other, they collide leading to a
merged object which is wildly oscillating and with a gravitational
mass which is above the maximum mass of the initial
configuration. Depending on whether the initial central rest-mass
density is larger or smaller than the critical one, the metastable
solution either collapses to a black hole (supercritical solutions) or
expands to a new stable stellar solution (subcritical
solutions). Exploiting the accuracy of the code and the large set of
initial configurations considered, we were able we are able to
identify the critical value for the central density with an accuracy
of 11 significant digits, a level of precision never achieved before
in this type of study.

Although overall more accurate, much of the results reported here
about the critical behaviour coincide and confirm those found by Jin
and Suen in~\cite{Jin:07a,wan_2008_das}. However, we do differ and
significantly when it comes to the interpretation of the critical
solutions. More specifically we have shown that the head-on collision
of two NSs near the critical threshold can be seen as series of
transitions in the space of configurations from an initial stable
model over to a metastable one with the same rest-mass but larger
gravitational mass as a result of the conversion of the kinetic energy
into internal energy via large shocks. The metastable solution is on
the linearly unstable branch of the equilibrium configurations and
thus it can exhibit a critical behaviour (see paper I) and either move
the stable branch of equilibria (subcritical solutions), or will
collapse to a black hole (supercritical solutions). Hence, the
critical solution is indeed a (perturbed) TOV solution, in contrast
with the conclusions drawn in~\cite{Jin:07a,wan_2008_das}

Finally, we have also computed the critical exponent of the scaling
relation of type-I critical phenomena and found it to agree well with
the one computed by~\cite{Jin:07a}. However, we have also found that,
superimposed with the standard power-law, the scaling law shows a fine
structure in terms of a periodic fluctuation. These wiggles in the
critical exponent have already been observed in the case of type-II
critical collapse of massless scalar fields~\cite{Gundlach97f, Hod97},
but were never reported before in the case of perfect fluids.

\ack It is a pleasure to thank Bruno Giacomazzo and Filippo Galeazzi
for their help and assistance with the \texttt{Whisky2D} code.  We are
also grateful to Shin'ichiro Yoshida for useful discussions. This work
was supported in part by the IMPRS on ``Gravitational-Wave
Astronomy'', by the DFG grant SFB/Transregio~7, and by ``CompStar'', a
Research Networking Programme of the European Science Foundation. The
computations were performed on the Damiana cluster at the AEI.

\appendix

\section{Metastable solutions and the hoop conjecture}
\begin{figure}
  \begin{center}
  \includegraphics[width=6.25cm]{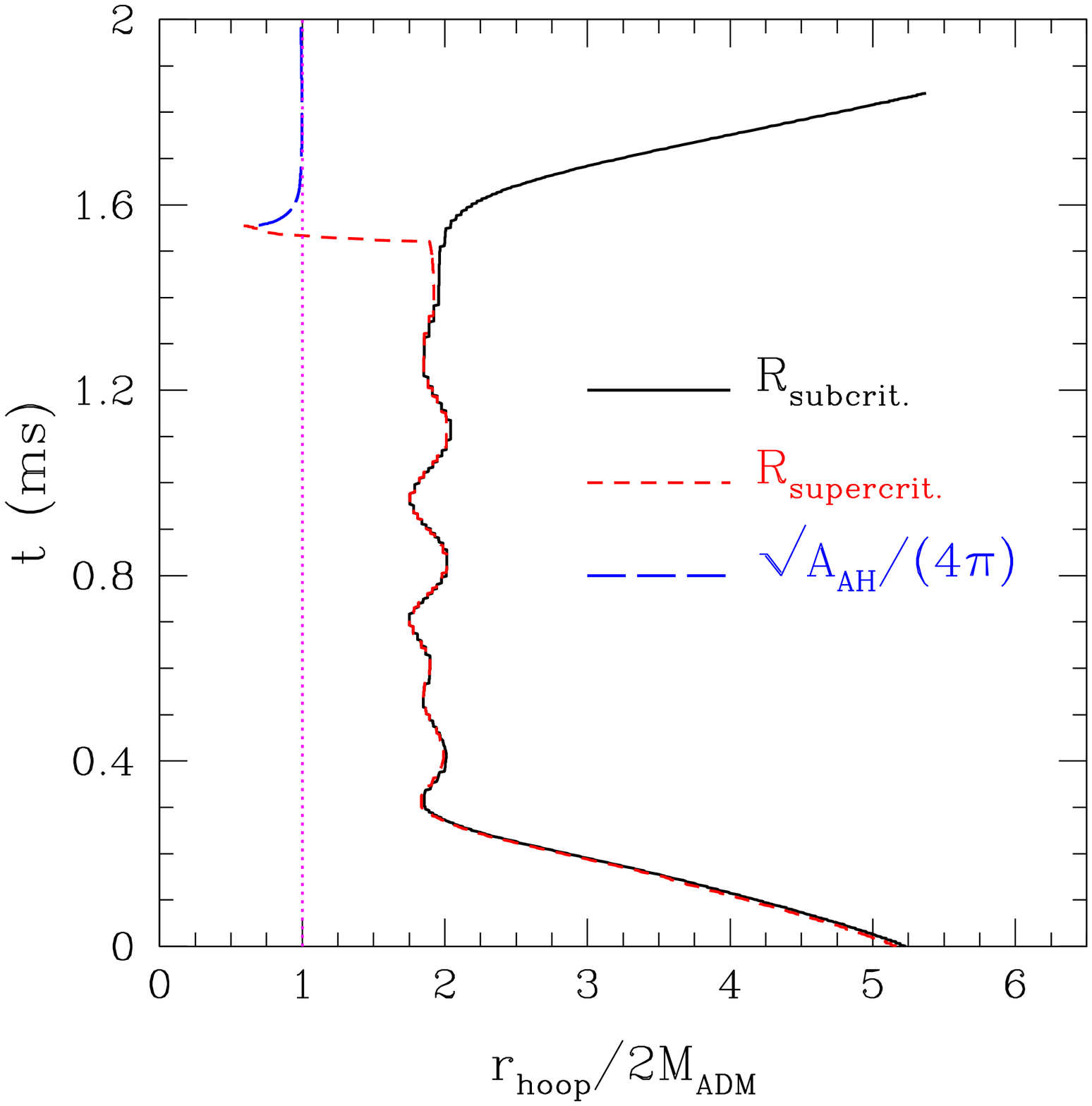}
\end{center}
  \vskip -0.5cm
  \caption{\label{fig:3} Spacetime diagram showing the worldines of
    the proper stellar radius for either a subcritical solution $R_{\rm
      subcrit.}$ (black solid line) or for a supercritical solution
    $R_{\rm supercrit.}$ (red dashed line). Both quantities are
    expressed in units of twice the ADM mass, so that the vertical
    dotted line at $1$ represents the threshold below which the hoop
    conjecture would be violated (the time coordinate is the proper
    time of an observer comoving at $R$). Also indicated with a blue
    long-dashed line is the worldline of the apparent horizon which is
    relevant in the case of the supercritical solution. Note that the
    metastable solution is always outside its ``hoop''.}
\end{figure}

As discussed in the main text, near the critical solution and during
the metastable stage, the evolution of all the hydrodynamical and
field variables is essentially the same (to a precision which is
proportional to the distance from the critical solution) for both
subcritical and supercritical solutions. It is therefore natural to
ask whether the metastable solution, which can either produce a black
hole or an expanded star, is ever compact enough to violate the hoop
conjecture. We recall that the conjecture states that a black hole is
formed if and only if a ``mass'' $M$ of matter is confined in a hoop
of radius which is in every direction is smaller or equal to the
corresponding Schwarzschild radius. Stated differently, a black hole
is expected to form if the matter is enclosed in a hoop of proper
circumference $\mathcal{C} \leq 4 \pi M$.

The hoop conjecture, as it was originally formulated, is not meant to
be a precise mathematical statement~\cite{wald_1997_gcc}, even though
it can be made precise under particular
circumstances~\cite{Flanagan91a, senovilla_2008_rhc,
  schoen_1983_ebh}. Most importantly, it is meant mostly as a
qualitative description of the gravitational collapse due to the
compression of matter. In particular the conjecture leaves much
freedom in the definition of both $\mathcal{C}$ and $M$. In view of
this freedom, we have taken $\mathcal{C}$ to be the proper
circumference of a coordinate circle enclosing $95\ \%$ of the mass of
the system (in order to exclude the extended low-density regions) and
used the ADM mass within $\mathcal{C}$ as a local measure of the mass
(This use of the ADM mass is strictly speaking incorrect as such a
mass is well defined only at spatial infinity; however it provides an
approximation which is adequate for the quality of the arguments made
here).

In figure \ref{fig:3} we show in a spacetime diagram the worldines of
the proper stellar radius for either a subcritical solution $R_{\rm
  subcrit.}$ (black solid line) or for a supercritical solution $R_{\rm
  supercrit.}$ (red dashed line). Both quantities are expressed in
units of twice the ADM mass, so that the vertical dotted line at $1$
represents the threshold below which the hoop conjecture would be
violated (the time coordinate is the proper time of an observer
comoving at $R$). Also indicated with a blue long-dashed line is the
worldline of the apparent horizon which is relevant in the case of the
supercritical solution. Quite clearly the metastable solution is
always outside its ``hoop'' (indeed about twice as large) and when the
supercritical solution crosses it is to produce a black hole as the
conjecture predicts. It remains unclear how these worldlines would
evolve if we had considered stars with larger boosts or that are are
far from the critical solution. These questions will be addressed in
our future work on the subject.

As a final remark we note that if we had used the interpretation of
the head-on collision as a transition between a stable and a
metastable TOV solution (\ie the one summarized in
figure~\ref{fig:2}), then it would have been rather obvious that the
hoop conjecture cannot be violated: a TOV star has a surface which is
always outside of its Schwarzschild radius.

\section*{References}
\bibliographystyle{iopart-num}
\bibliography{head_on_collision}
\end{document}